\documentclass[useAMS,usegraphicx,usenatbib]{mn2e}

\newcommand{\Msun}{\ensuremath{\,{\rm M}_\odot}}           
\newcommand{\kms}{\,km\,s$^{-1}$}                          
\newcommand{\Apx}{\,\AA\,px$^{-1}$}                        
\newcommand{\apx}{$^{\prime\prime}$\,px$^{-1}$}            
\newcommand{\ion}[2]{{#1}\,{\sc {\small{#2}}}}             
\newcommand{\Porb}{\ensuremath{P_{\rm orb}}}               
\newcommand{\mc}[1]{\multicolumn{2}{c}{#1}}
\newcommand{\cd}{\ensuremath{\,{\rm cycle\,\,d}^{-1}}}     
\newcommand{\as}{\ensuremath{^{\prime\prime}}}             
\newcommand{\reff}[1]{{#1}}                                  

\title[Orbital periods of SDSS cataclysmic variables. V.]
      {Orbital periods of cataclysmic variables identified by the SDSS. V.
       VLT, NTT and Magellan observations of nine equatorial systems}

\author[Southworth et al.]
       {John Southworth$^1$%
        \thanks{E-mail: j.k.taylor@warwick.ac.uk (JS),
        \newline Boris.Gaensicke@warwick.ac.uk (BTG),
        \newline T.R.Marsh@warwick.ac.uk (TRM)},
        B.\ T.\ G\"ansicke$^1$,
        T.\ R.\ Marsh$^1$,
        M.\ A.\ P.\ Torres$^2$,
        D.\ Steeghs$^{1,2}$,
        \newauthor
        P.\ Hakala$^3$,
        C.\ M.\ Copperwheat$^1$,
        A.\ Aungwerojwit$^{1,4}$,
        A.\ Mukadam$^5$
        \\
        $^1$ Department of Physics, University of Warwick, Coventry, CV4 7AL, UK \\
        $^2$ Harvard‐Smithsonian Center for Astrophysics, 60 Garden Street, Cambridge, MA 02138, USA  \\
        $^3$ Tuorla Observatory, University of Turku, FIN-21500 Piikki\"o, Finland \\
        $^4$ Department of Physics, Faculty of Science, Naresuan University, Phitsanulok, 65000, Thailand \\
        $^5$ Department of Astronomy, University of Washington, Box 351580, Seattle, WA 98195, USA
        }

\begin{document} \maketitle 

\begin{abstract}
We present VLT and Magellan spectroscopy and NTT photometry of nine faint cataclysmic variables (CVs) which were spectroscopically identified by the Sloan Digital Sky Survey. We measure orbital periods for five of these from the velocity variations of the cores and wings of their H$\alpha$ emission lines. Four of the five have orbital periods shorter than the 2--3\,hour period gap observed in the known population of CVs. SDSS J004335.14$-$003729.8 has an orbital period of $\Porb = 82.325 \pm 0.088$\,min; Doppler maps show emission from the accretion disc, bright spot and the irradiated inner face of the secondary star. In its light curve we find a periodicity which may be attributable to pulsations of the white dwarf. SDSS J163722.21$-$001957.1 has $\Porb = 99.75 \pm 0.86$\,min. By combining this new measurement with a published superhump period we estimate a mass ratio of $q \approx 0.16$ and infer the physical properties and orbital inclination of the system. For SDSS J164248.52$+$134751.4 we find $\Porb = 113.60 \pm 1.5$\,min. The Doppler map of this CV shows an unusual brightness distribution in the accretion disc which would benefit from further observations. SDSS J165837.70$+$184727.4 had spectroscopic characteristics which were very different between the SDSS spectrum and our own VLT observations, despite only a small change in brightness. We measure $\Porb = 98.012 \pm 0.065$\,min from its narrow H$\alpha$ emission line. Finally, SDSS J223843.84$+$010820.7 has a comparatively longer period of $\Porb = 194.30 \pm 0.16$\,min. It contains a magnetic white dwarf and, with $g = 18.15$, is brighter than the other objects studied here. These results continue the trend for the fainter CVs identified by the SDSS to be almost exclusively shorter-period objects with low mass transfer rates.
\end{abstract}



\section{INTRODUCTION}                                                                       \label{sec:intro}

Cataclysmic variables (CVs) are interacting binary stars containing a white dwarf primary component in a close orbit with a low-mass secondary star which fills its Roche lobe. In most of these systems the secondary component is hydrogen-rich and transfers material to the white dwarf via an accretion disc. Comprehensive reviews of the properties of CVs have been given by \citet{Warner95book} and \citet{Hellier01book}.

The evolution of CVs is thought to be governed primarily by the loss of orbital angular momentum due to gravitational radiation \citep{Paczynski67aca} and magnetic braking \citep{VerbuntZwaan81aa,Rappaport++82apj}. These effects \reff{are predicted to} cause CVs to evolve towards shorter orbital periods until a minimum value of about 80\,min, at which point the secondary stars become degenerate and the CVs evolve back to longer periods \reff{(e.g.\ \citealt{Patterson98pasp})}.

The distribution of orbital periods of the observed population of CVs has a characteristic `period gap' \citep{WhyteEggleton80mn,Knigge06mn} in the interval between 2.2 and 3.2 hours. The deficiency in the number of systems in this period range is thought to result from the sudden cessation of magnetic braking due to structural changes in CV secondary stars \citep{SpruitRitter83aa}. The number of CVs shortwards of this period gap are observed to be roughly equal to the number which are longward of the gap \citep[e.g.][]{Downes+01pasp,RitterKolb03aa}.

Unfortunately, theoretical studies of the population of CVs have consistently predicted that the vast majority of these objects should have short periods ($\Porb \la 2$\,hours) due to their relatively longer evolutionary timescale \citep{Dekool92aa, DekoolRitter93aa, Kolb93aa, KolbDekool93aa, Politano96apj, Politano04apj, KolbBaraffe99mn, Howell++01apj, Willems+05apj}, culminating in a strong `spike' in the population at a minimum period of about 65\,min. The remarkable differences between the predicted and observed populations of CVs have not yet been satisfactorily explained, although it is clear that observational selection biases have a lot to answer for.

To understand these selection biases, and to discover what the properties of the intrinsic population of CVs are, we are conducting a research program to characterise the sample of CVs identified by the Sloan Digital Sky Survey (SDSS\footnote{\tt http://www.sdss.org/}; \citealt{York+00aj}). A total of 212 of these objects have been found spectroscopically from an initial selection based on single-epoch photometric colour indices \citep{Szkody+02aj, Szkody+03aj, Szkody+04aj, Szkody+05aj, Szkody+06aj, Szkody+07aj}. This sample is therefore not biased towards CVs which are variable or strong X-ray emitters, and also has a much wider coverage of colour space than previous large-scale surveys \citep{Green++86apjs,Chen+01mn,Aungwerojwit+06aa}. Results and further discussion of our project to measure the orbital periods of SDSS CVs can be found in \citet{Gansicke+06mn}, \citet[][hereafter Paper\,I]{Me+06mn}, \citet{Me+07mn,Me+07mn2,Me++08mn}, \citet{Dillon+08mn,Dillon+08}, and \citet{Littlefair+06mn, Littlefair+06sci, Littlefair+07mn, Littlefair+08}.

\begin{figure} \includegraphics[width=0.48\textwidth,angle=0]{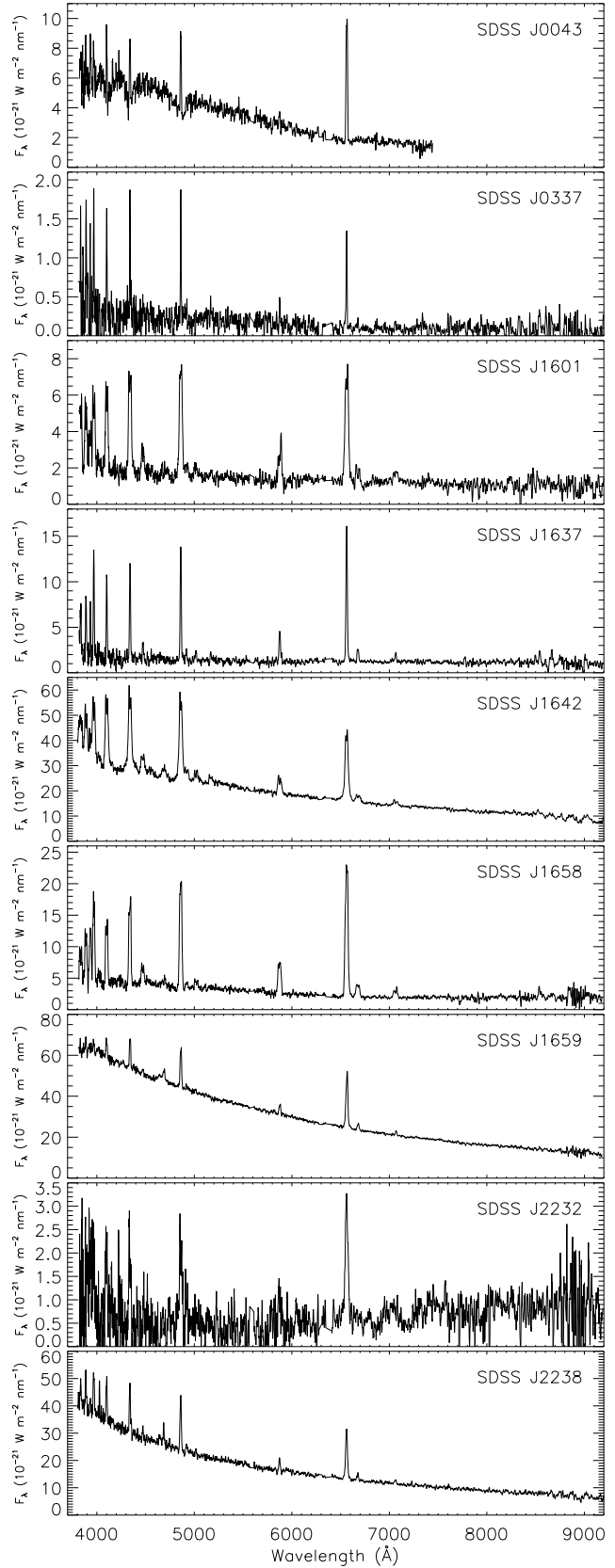} \\
\caption{\label{fig:sdssspec} SDSS spectra of the CVs studied in this work. For
this plot the flux levels have been smoothed with 10-pixel Savitsky-Golay filters.
The units of the abscissae are $10^{-21}$\,W\,m$^{-2}$\,nm$^{-1}$, which corresponds
to $10^{-17}$\,erg\,s$^{-1}$\,cm$^{-2}$\,\AA$^{-1}$.}\end{figure}

In this work we present time-resolved spectroscopy and photometry of nine CVs, and measure orbital period for five of these. We shall abbreviate the names of the targets to SDSS\,J0043, SDSS\,J0337, SDSS\,J1601, SDSS\,J1637, SDSS\,J1642, SDSS\,J1658, SDSS\,J1659, SDSS\,J2232 and SDSS\,J2238. Their full names and $ugriz$ apparent magnitudes are given in Table\,\ref{tab:iddata}.
In Fig.\,\ref{fig:sdssspec} we have plotted their SDSS spectra for reference.

\begin{table*} \begin{center}
\caption{\label{tab:iddata} Apparent magnitudes of our targets in the SDSS $ugriz$
passbands. $g_{\rm spec}$ are apparent magnitudes we have calculated by convolving
the SDSS flux-calibrated spectra with the $g$ passband function. They are obtained
at a different epoch to the $ugriz$ magnitudes measured from the imaging observations,
but are less reliable as they are affected by `slit losses', and any errors in astrometry
or positioning of the spectroscopic fibre entrance.}
\begin{tabular}{lllccccccc} \hline
SDSS name                  & Short name  & Reference           & $u$ & $g$ & $r$ & $i$ & $z$ & $g_{\rm spec}$  \\
\hline
SDSS J004335.14$-$003729.8 & SDSS\,J0043 & \citet{Szkody+04aj} & 20.18 & 19.86 & 19.81 & 19.97 & 19.84 & 19.95 \\
SDSS J033710.91$-$065059.4 & SDSS\,J0337 & \citet{Szkody+07aj} & 19.64 & 19.54 & 19.72 & 19.97 & 20.14 & 23.27 \\
SDSS J160111.53$+$091712.6 & SDSS\,J1601 & \citet{Szkody+06aj} & 19.96 & 20.11 & 20.12 & 20.22 & 19.74 & 20.39 \\
SDSS J163722.21$-$001957.1 & SDSS\,J1637 & \citet{Szkody+02aj} & 16.83 & 16.60 & 16.59 & 16.75 & 16.84 & 20.57 \\
SDSS J164248.52$+$134751.4 & SDSS\,J1642 & \citet{Szkody+07aj} & 18.48 & 18.64 & 18.50 & 18.42 & 18.21 & 18.03 \\
SDSS J165837.70$+$184727.4 & SDSS\,J1658 & \citet{Szkody+06aj} & 20.51 & 20.07 & 20.12 & 20.09 & 19.68 & 19.71 \\
SDSS J165951.68$+$192745.6 & SDSS\,J1659 & \citet{Szkody+06aj} & 16.84 & 16.73 & 16.78 & 16.86 & 16.96 & 17.12 \\
SDSS J223252.35$+$140353.0 & SDSS\,J2232 & \citet{Szkody+04aj} & 17.74 & 17.66 & 17.80 & 17.88 & 17.97 & 23.16 \\
SDSS J223843.84$+$010820.7 & SDSS\,J2238 & \citet{Szkody+03aj} & 18.29 & 18.15 & 18.08 & 18.18 & 18.17 & 18.27 \\
\hline \end{tabular} \end{center} \end{table*}


\begin{table*} \begin{center}
\caption{\label{tab:obslog} Log of the observations presented in this work.
The acquisition magnitudes were measured from the VLT/FORS2 acquisition images
and are discussed in Section\,\ref{sec:obs:vltphot}. The passbands for the
acquisition magnitudes are indicated, where `$V$' denotes the Johnson $V$ band
and `$Wh$' indicates that no filter was used.}
\begin{tabular}{lcccccccc} \hline
Target    & Date & Start time & End time & Telescope and      & Optical         &  Number of   & Exposure & Mean          \\
          & (UT) &  (UT)      &  (UT)    &  instrument        & element         & observations & time (s) & magnitude     \\
\hline
SDSS\,J0043 & 2007 08 16 & 07:47 & 10:25 & VLT\,/\,FORS2      & 1200R grism                  &  21 &   400  & $V = 19.9$  \\
SDSS\,J0043 & 2007 08 17 & 07:13 & 09:03 & VLT\,/\,FORS2      & 1200R grism                  &  14 &   440  & $V = 19.8$  \\
SDSS\,J0043 & 2005 10 08 & 05:05 & 06:49 & APO\,3.5m\,/\,DIS  & unfiltered                   & 175 & 15--20 & $V = 19.8$  \\
[2pt]
SDSS\,J0337 & 2007 08 17 & 09:20 & 10:18 & VLT\,/\,FORS2      & 1200R grism                  &   6 &   600  & $V = 21.4$  \\
[2pt]
SDSS\,J1601 & 2007 08 10 & 23:25 & 00:10 & NTT\,/\,SUSI2      & unfiltered                   &  36 &    60  & $Wh = 20.1$ \\
[2pt]
SDSS\,J1637 & 2007 08 16 & 01:18 & 04:06 & VLT\,/\,FORS2      & 1200R grism                  &  20 &600--400& $V = 20.3$  \\
SDSS\,J1637 & 2007 08 16 & 23:28 & 00:40 & VLT\,/\,FORS2      & 1200R grism                  &   8 &   480  & $V = 20.6$  \\
[2pt]
SDSS\,J1642 & 2007 08 06 & 23:38 & 02:55 & NTT\,/\,SUSI2      & $V$ filter                   & 262 &    30  & $V = 19.5$  \\
SDSS\,J1642 & 2007 08 07 & 23:21 & 04:21 & NTT\,/\,SUSI2      & $V$ filter                   & 414 &    28  & $V = 19.6$  \\
SDSS\,J1642 & 2007 08 17 & 00:48 & 03:41 & VLT\,/\,FORS2      & 1200R grism                  &  29 &   300  & $V = 18.5$  \\
[2pt]
SDSS\,J1658 & 2007 08 15 & 00:17 & 03:23 & VLT\,/\,FORS2      & 1200R grism                  &  19 &   480  & $Wh = 19.5$ \\
SDSS\,J1658 & 2007 08 15 & 23:33 & 01:01 & VLT\,/\,FORS2      & 1200R grism                  &  12 &   400  & $V = 20.1$  \\
[2pt]
SDSS\,J1659 & 2007 08 11 & 00:19 & 02:45 & NTT\,/\,SUSI2      & $V$ filter                   & 166 & 30--60 & $V = 17.0$  \\
[2pt]
SDSS\,J2232 & 2005 08 11 & 00:10 & 05:34 & NOT\,/\,ALFOSC     & unfiltered                   & 177 &    60  & $Wh = 22.1$ \\
SDSS\,J2232 & 2006 07 21 & 02:03 & 05:12 & NOT\,/\,ALFOSC     & unfiltered                   & 109 &    60  & $Wh = 21.8$ \\
SDSS\,J2232 & 2007 08 15 & 03:37 & 06:10 & VLT\,/\,FORS2      & 1200R grism                  &  14 &   600  & $V = 21.4$  \\
[2pt]
SDSS\,J2238 & 2007 08 14 & 07:27 & 09:44 & Magellan\,/\,IMACS & 600\,$\ell$\,mm$^{-1}$ grism &  22 &   300  &             \\
SDSS\,J2238 & 2007 08 15 & 07:24 & 09:51 & Magellan\,/\,IMACS & 600\,$\ell$\,mm$^{-1}$ grism &  17 &   300  &             \\
SDSS\,J2238 & 2007 08 16 & 07:39 & 09:42 & Magellan\,/\,IMACS & 600\,$\ell$\,mm$^{-1}$ grism &  19 &   300  &             \\
\hline \end{tabular} \end{center} \end{table*}

\section{OBSERVATIONS AND DATA REDUCTION}                                           \label{sec:obs}

\reff{A log of observations is given in Table\,\ref{tab:obslog}. The reduced data and velocity measurements presented here will be made available at the CDS ({\tt http://cdsweb.u-strasbg.fr/}) and at {\tt http://www.astro.keele.ac.uk/$\sim$jkt/}.}

\subsection{VLT spectroscopy}                                                       \label{sec:obs:vltspec}

Spectroscopic observations were obtained in 2007 August using the FORS2 spectrograph \citep{Appenzeller+98msngr} and Very Large Telescope (VLT) at ESO Paranal, Chile. The 1200R grism was used for all observations, giving a wavelength interval of 5870\,\AA\ to 7370\,\AA\ with a reciprocal dispersion of 0.73\Apx. From measurements of the full widths at half maximum (FWHMs) of arc-lamp and night-sky spectral emission lines, we estimate a resolution of 1.6\,\AA\ at H$\alpha$.

The data were reduced using optimal extraction (\citealt{Horne86pasp}) as implemented in the {\sc pamela}\footnote{{\sc pamela} and {\sc molly} were written by TRM and can be found at \\ {\tt http://www.warwick.ac.uk/go/trmarsh}} code (\citealt{Marsh89pasp}), and the {\sc starlink}\footnote{Starlink software can be accessed from \\ {\tt http://starlink.jach.hawaii.edu/}} packages {\sc figaro} and {\sc kappa}.

The wavelength calibration of the spectra was obtained using one arc spectrum per night, taken during daytime as part of the standard calibration routines for FORS2. As in Paper\,I we have found that flexure of the spectrograph can cause night-time spectra to shift by up to 42\kms\ (1.2 pixels) depending on elevation. We have removed this trend from each spectrum by measuring the position of the \ion{O}{I} night sky emission line at 6300.304\,\AA\ \citep{Osterbrock+96pasp}. Offsets calculated using other lines (e.g.\ \ion{O}{I} 6363.78\,\AA) always agree to within 0.1 pixels.

\begin{figure*} \includegraphics[width=\textwidth,angle=0]{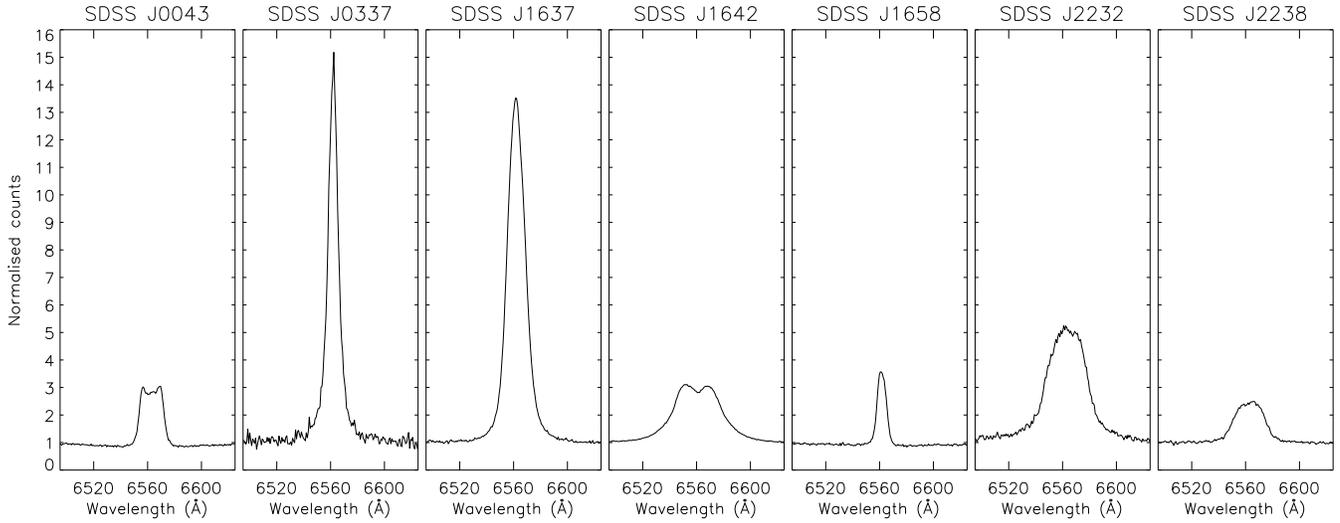} \\
\caption{\label{fig:Halpha} The averaged H$\alpha$ emission line profiles of
the seven CVs for which we present spectroscopy. The spectra have had their
continuum level normalised to unity. The orbital motion was removed from the
spectra of SDSS\,J1658 before its mean spectrum was constructed.} \end{figure*}

The VLT spectra cover the H$\alpha$ and \ion{He}{I} 6678 and 7065\,\AA\ lines. The average continuum-normalised H$\alpha$ profile for each object is plotted in Fig.\,\ref{fig:Halpha}.

\begin{figure*}        
\includegraphics[width=0.19\textwidth,angle=0]{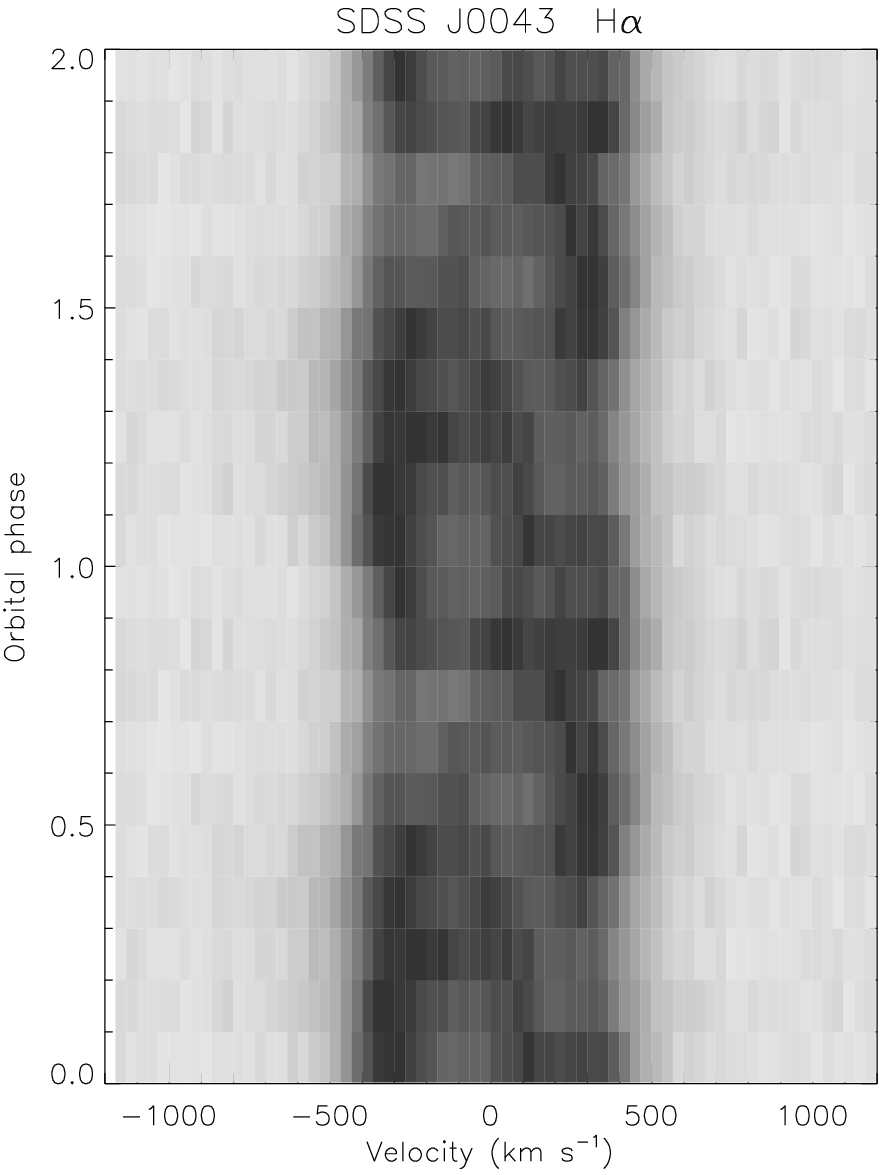}
\includegraphics[width=0.19\textwidth,angle=0]{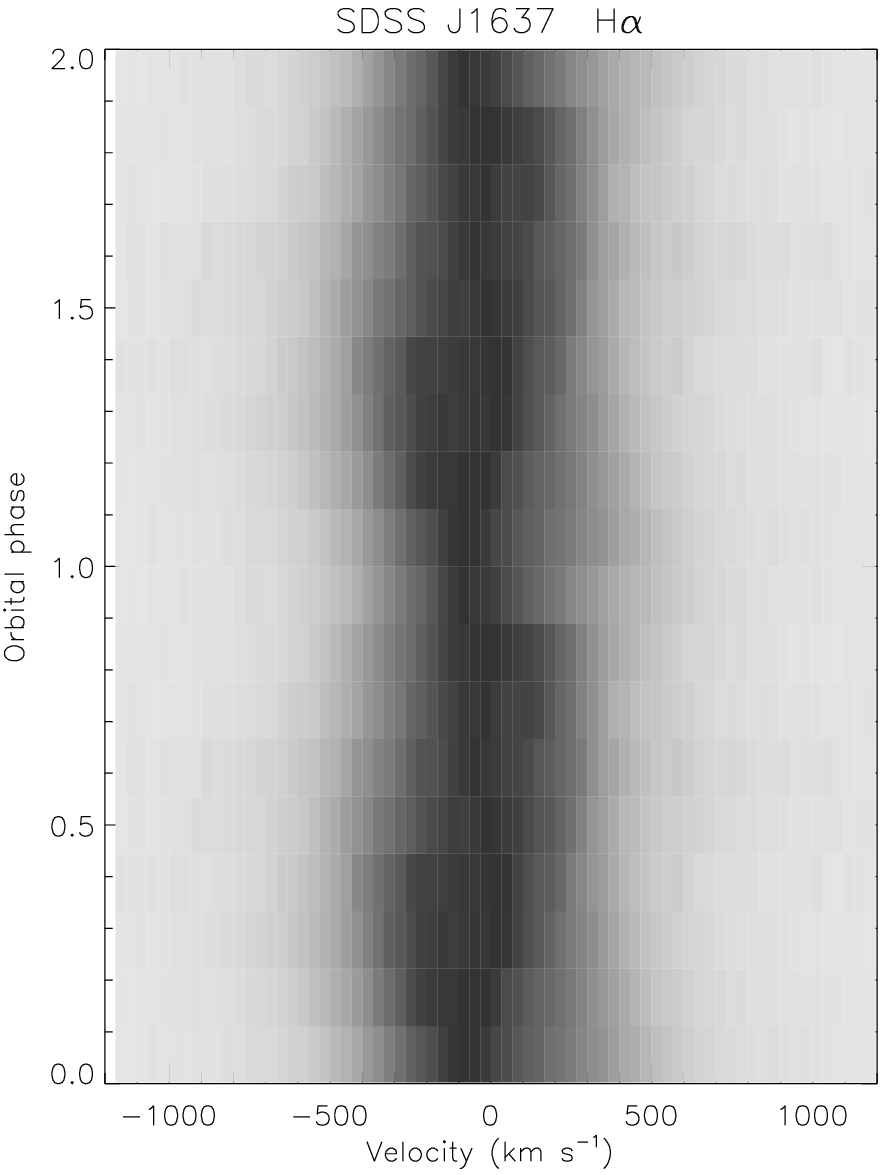}
\includegraphics[width=0.19\textwidth,angle=0]{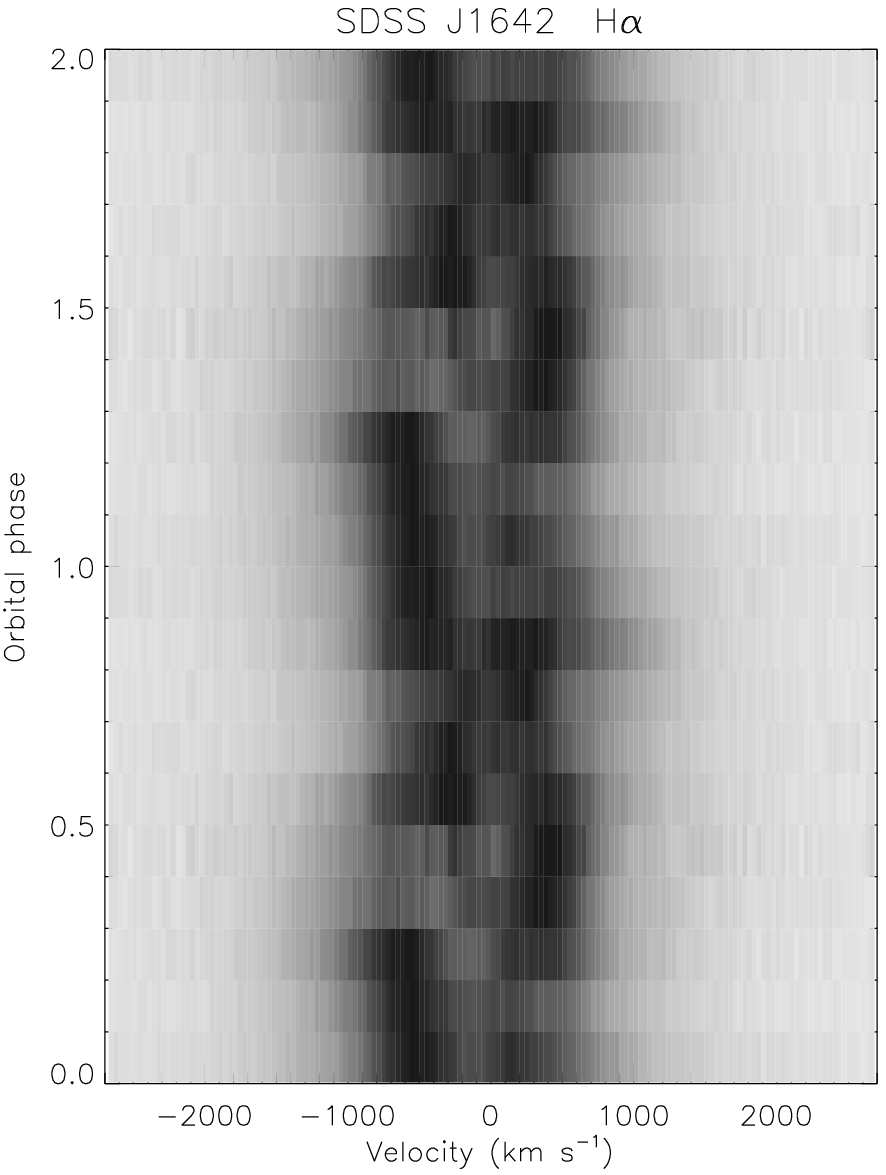}
\includegraphics[width=0.19\textwidth,angle=0]{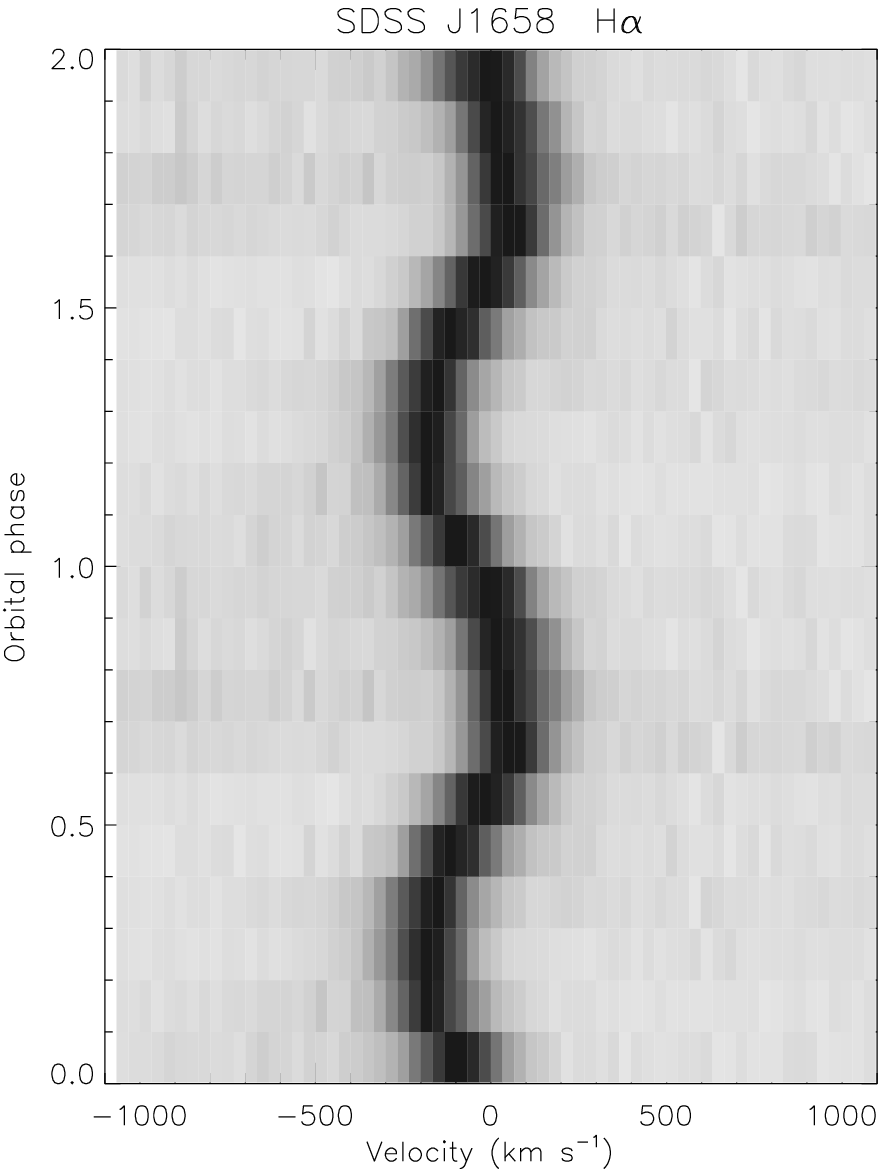}
\includegraphics[width=0.19\textwidth,angle=0]{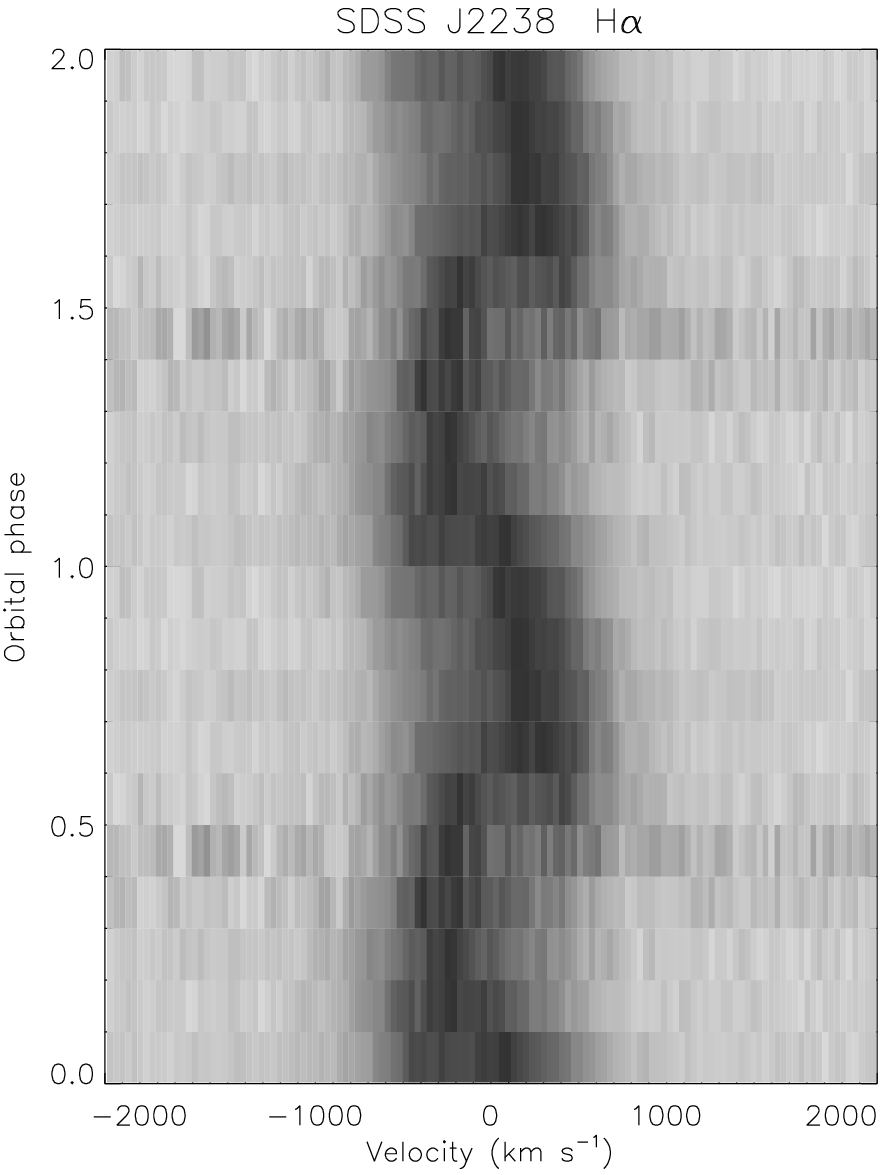}
\includegraphics[width=0.19\textwidth,angle=0]{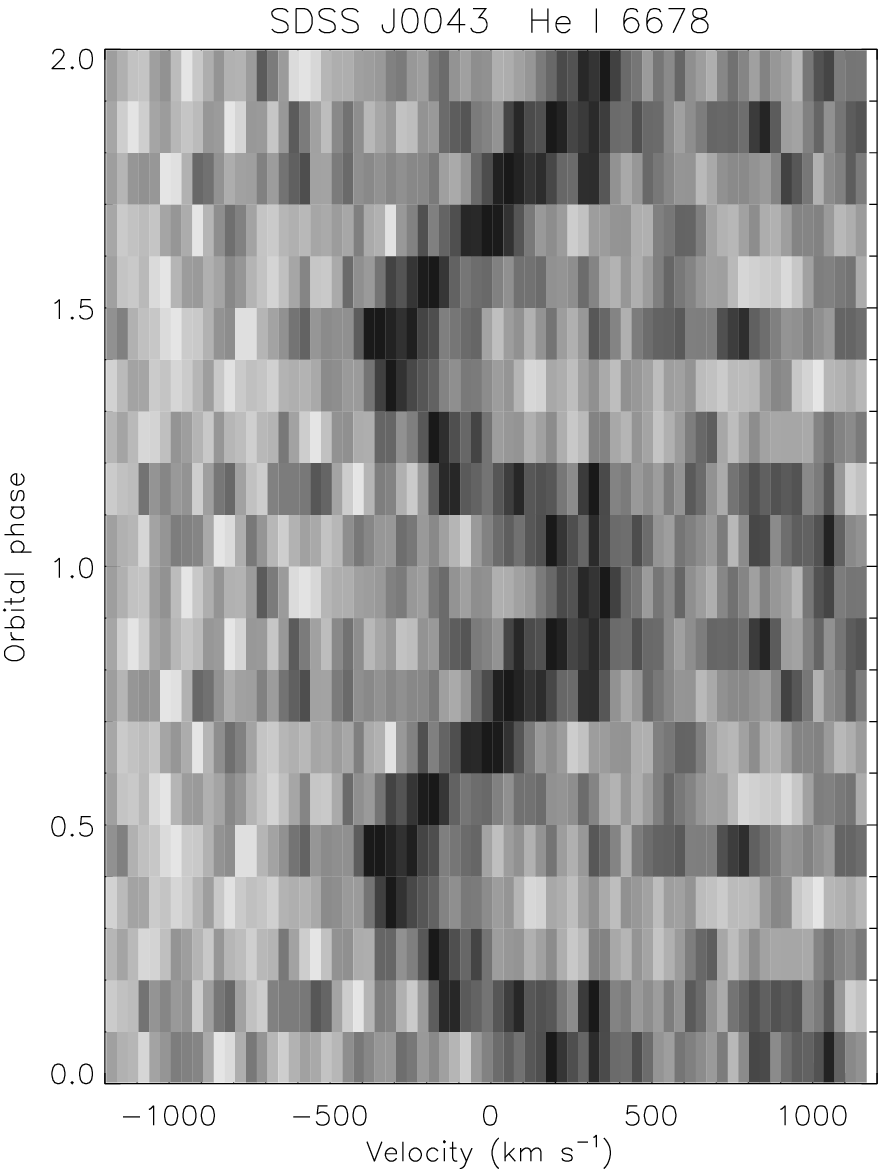}
\includegraphics[width=0.19\textwidth,angle=0]{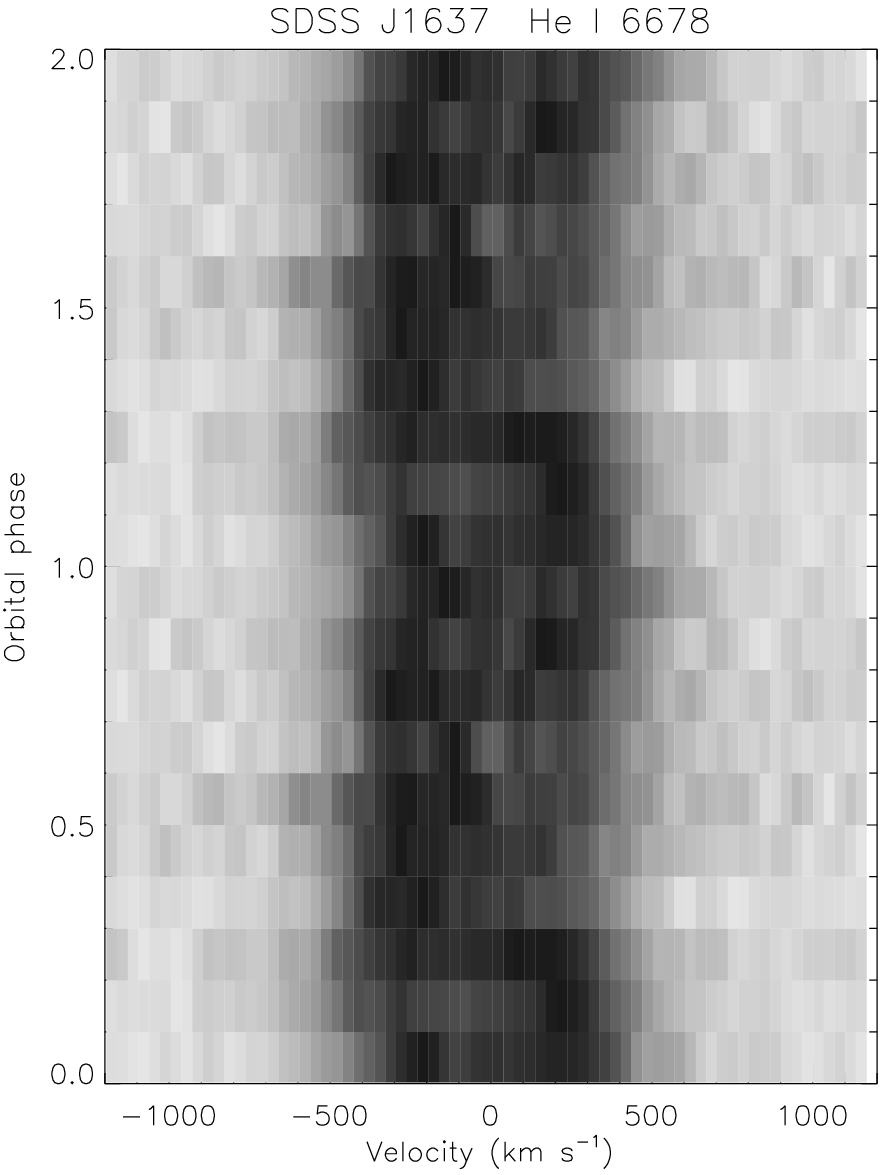}
\includegraphics[width=0.19\textwidth,angle=0]{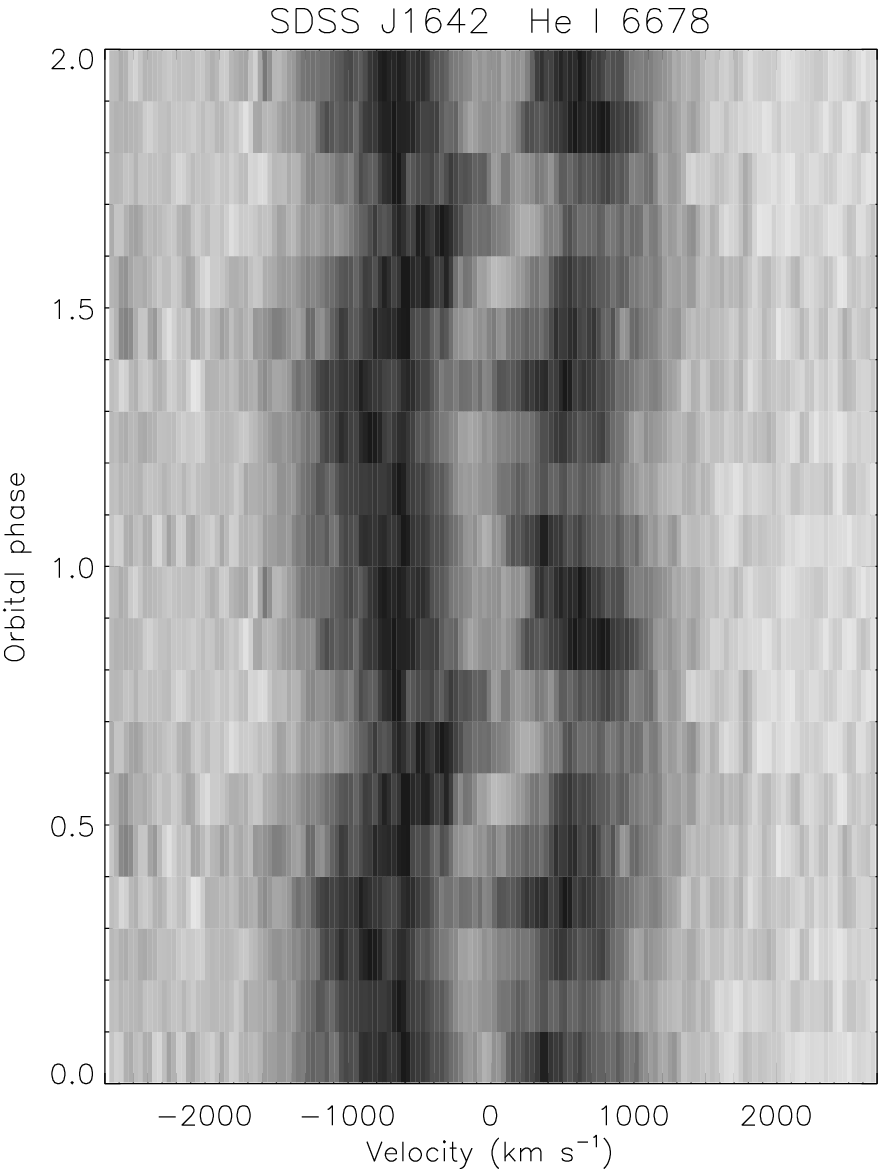}
\includegraphics[width=0.19\textwidth,angle=0]{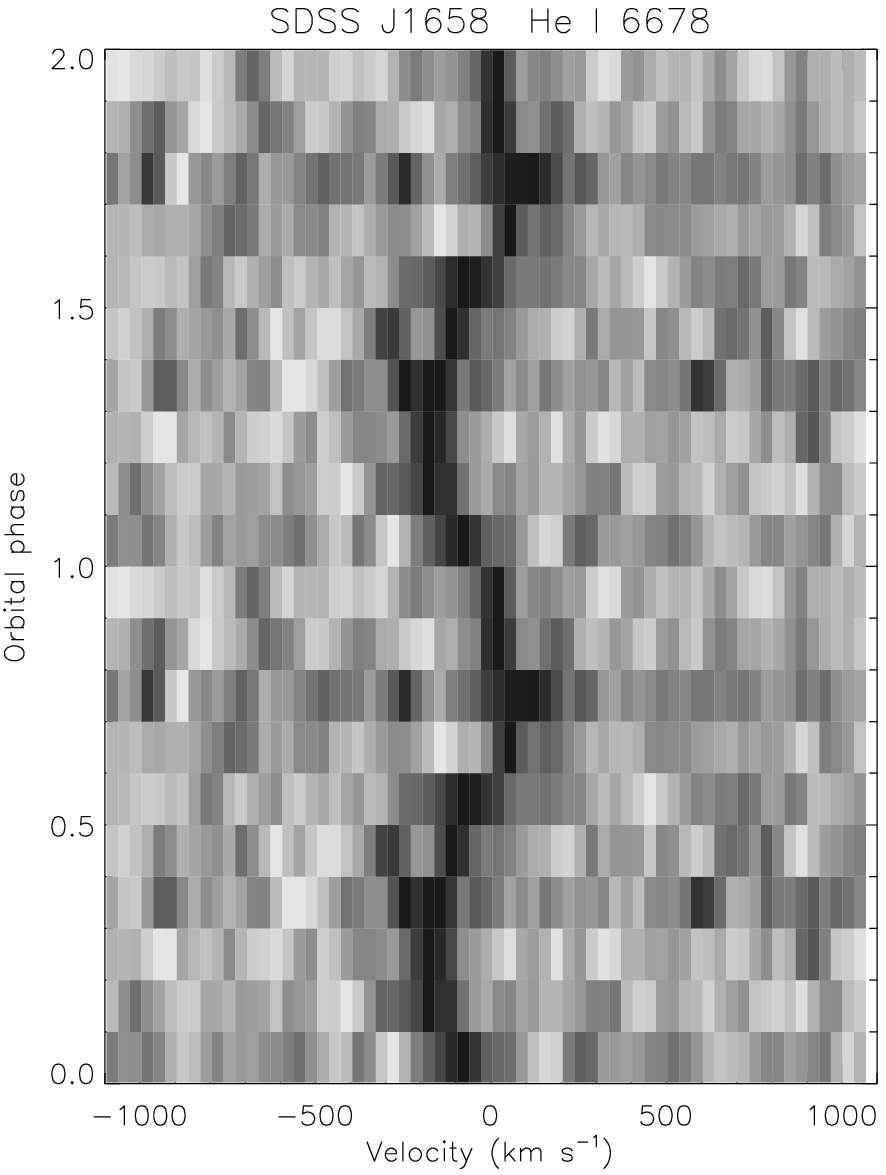}
\includegraphics[width=0.19\textwidth,angle=0]{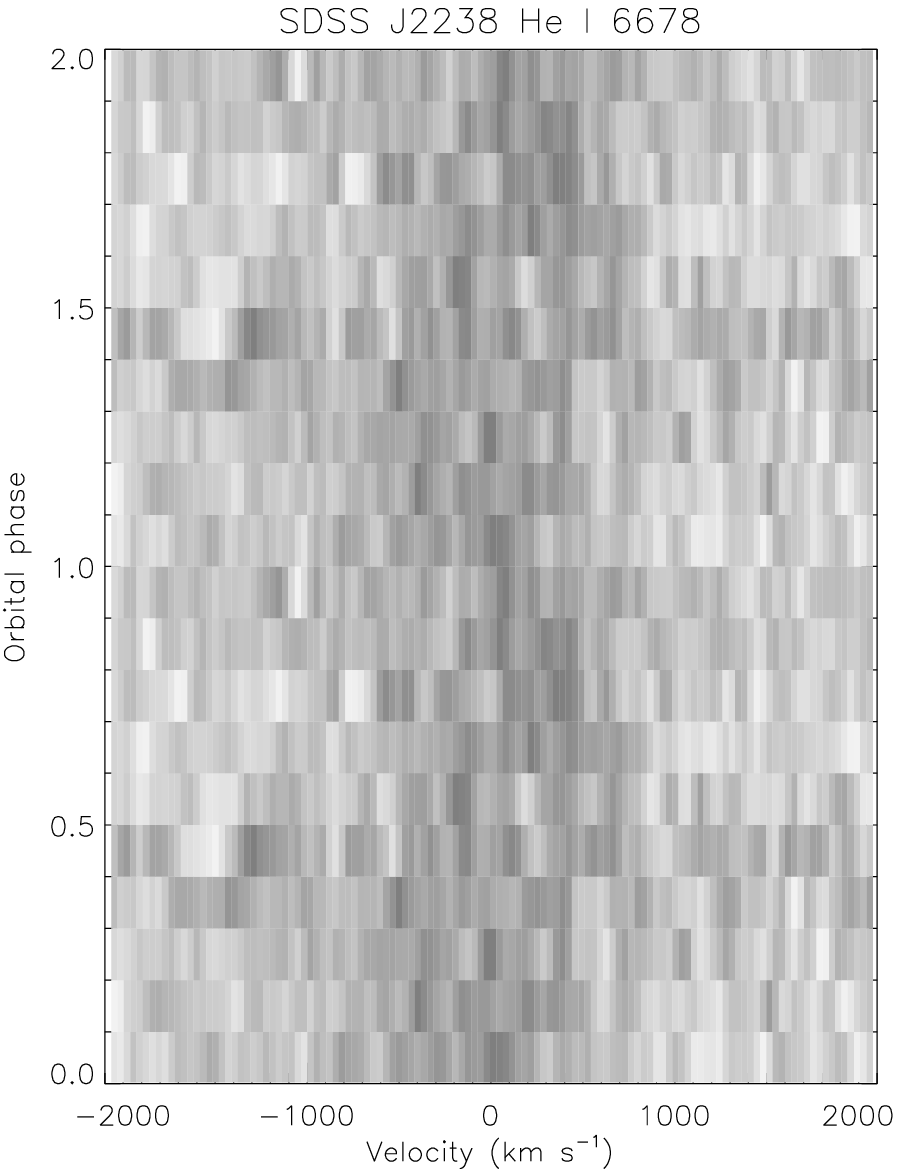}
\caption{\label{fig:vlt:trailed} Greyscale plots of the continuum-%
normalised and phase-binned trailed spectra of the five CVs for which
we obtained orbital periods. Darker shading indicates stronger emission.
The upper plots show the H$\alpha$ lines. The lower plots show the
\ion{He}{I} 6678\,\AA\ lines, with the same velocity scale but different
intensity scale. The \ion{He}{I} spectra have been smoothed with a
Savitsky-Golay filter for display purposes.} \end{figure*}

\subsection{VLT photometry}                                                               \label{sec:obs:vltphot}

The observing procedure of FORS2 included obtaining target acquisition images. Exposure times were generally 20\,s and the observations used either Johnson $V$ or were unfiltered. We have extracted differential photometry from these images using the {\sc starlink} package {\sc gaia}, taking special care to select comparison stars from the SDSS database with colours as close to our target CVs as possible in order to minimise colour effects. The $V$-band apparent magnitudes of the comparison stars were calculated from the $g$ and $r$ magnitudes using the transformations provided by \citet{Jester+05aj}.

\subsection{Magellan spectroscopy}                                                  \label{sec:obs:mag}

Time-resolved spectroscopy of SDSS\,J2238 was obtained in 2007 August using the Inamori-Magellan Areal Camera and Spectrograph (IMACS; \citealt{BigelowDressler03spie}) on the 6.5m Magellan Baade Telescope at Las Campanas Observatory (Table\,\ref{tab:obslog}). IMACS was employed in long-camera mode, using a 600 line\,mm$^{-1}$ grating centred at 5550\,\AA. This instrumental setup yielded a reciprocal dispersion of 0.76\Apx\ in the spectral interval 3920--7100\,\AA. The spectra were obtained dispersed along the short axis of four of the eight SITe CCDs in the IMACS mosaic detector. On August 14, most spectra were obtained with a 0.7 arcsec slit width which yielded a spectral resolution of 1.7\,px (1.3\,\AA). The other data were obtained with a 1.2 arcsec slit width that provided a spectral resolution of 3.5\,px (2.7\,\AA). Weather conditions during were generally good, with seeing ranging from 0.7\as\ to 1.2\as\ FWHM.

The IMACS frames were bias and flat-field corrected with standard IRAF routines. The spectra were extracted from each CCD frame with the IRAF {\sc kpnoslit} package. The wavelength calibration was derived from cubic spline fits to HeNeAr lamp spectra bracketing the target spectra. The root mean square deviation of the fits was always less than 0.05\,\AA.

\subsection{NTT photometry}                                                               \label{sec:obs:ntt}

SDSS\,J1601, SDSS\,J1642 and SDSS\,J1659 were observed using the New Technology Telescope (NTT) at ESO La Silla, Chile. Time-series imaging photometry was obtained using the SUSI2 high-resolution imager \citep{Dodorico+98spie}. The observing run was affected by cloud, wind, snow and ice. For SDSS\,J1642 and SDSS\,J1659 we used a Johnson $V$ filter. For SDSS\,J1601 we obtained unfiltered photometry and took care to avoid regions of the CCD where fringing was strong. The CCD was binned by factors of three in both directions, giving a spatial resolution of 0.24\apx.

Debiasing and flat-fielding of the raw images was performed with the {\sc starlink} software packages {\sc convert} and {\sc kappa}. Optimal and aperture differential photometry \citep{Naylor98mn} was measured from the reduced images with the {\sc multiphotom} script \citep{Me++04mn}, which uses the {\sc autophotom} package \citep{Eaton++99} to obtain time-series photometry. Differential magnitudes were converted into apparent mangitudes in way described in Section.\,\ref{sec:obs:vltphot}.

\subsection{NOT photometry}                                                               \label{sec:obs:not}

Light curves of SDSS\,J2232 were obtained in 2005 August and 2006 July using the Nordic Optical Telescope (NOT) and ALFOSC imaging spectrograph. The observations were unfiltered, windowed, and binned by a factor of 2 in both directions. The detector was an EEV 2k$\times$4k pixel CCD with an unbinned plate scale of 0.19\as\,px$^{-1}$.

These data were reduced using the pipeline described by \citet{gansicke+04aa}, which performs bias and flat-field corrections within {\sc midas}\footnote{\tt http://www.eso.org/projects/esomidas/} and uses the {\sc sextractor} package \citep{BertinArnouts96aas} to perform aperture photometry for all objects in the field of view. Differential magnitudes were converted into apparent magnitude in the same way as for our NTT and VLT photometry.

\subsection{APO photometry}                                                               \label{sec:obs:apo}

A light curve of SDSS\,J0043 was observed on 2005 October 8th using the 3.5\,m telescope at Apache Peak Observatory (APO) and Dual Imaging Spectrograph (DIS) in imaging mode. With this instrument the blue and red portions of the beam are split by a dichroic at 5550\,\AA. We present here the blue-arm photometry, which is unfiltered but sensitive to light in the wavelength interval 3500--5550\,\AA. The CCD had a plate scale of 0.4\Apx, and windowing was used to decrease the readout time. We used a standard IRAF reduction to extract sky-subtracted light curves from the CCD frames using weighted circular aperture photometry \citep{Odonoghue+00balta}.


\section{DATA ANALYSIS}

\begin{table*} \begin{center}
\caption{\label{tab:orbits} Circular spectroscopic orbits found using {\sc sbop}.
Phase zero corresponds to the blue-to-red crossing point of the RVs; because we are
measuring emission lines from the accretion disc this usually has a phase offset with
respect to inferior conjunction of the white dwarf (see Section\,\ref{sec:1642}).}
\begin{tabular}{l r@{\,$\pm$\,}l r@{\,$\pm$\,}l r@{\,$\pm$\,}l r@{\,$\pm$\,}l r}\hline
Target&\mc{Orbital period}&\mc{Reference time}&\mc{Velocity amplitude}&\mc{Systemic velocity}&$\sigma_{\rm rms}$\\
      &     \mc{(day)}    &    \mc{(HJD)}     &       \mc{(\kms)}     &      \mc{(\kms)}     &      (\kms)      \\
\hline 
SDSS\,J0043 &  0.0571702 & 0.000061  &  2454328.87140 & 0.00071  &            50.0 & 2.8  &    13.1 & 2.0  & 11.8 \\
SDSS\,J1637 &  0.067391  & 0.00013   &  2454326.6720  & 0.0043   &            24.4 & 1.6  &  $-$9.1 & 1.2  &  6.0 \\
SDSS\,J1642 &  0.07889   & 0.0011    &  2454329.55238 & 0.00074  &           105.9 & 2.4  & $-$55.1 & 2.4  & 12.7 \\
SDSS\,J1658 &  0.0680638 & 0.000045  &  2454327.54703 & 0.00037  &           125.3 & 3.5  & $-$41.5 & 2.4  & 13.5 \\
SDSS\,J2238 &  0.134932  & 0.00011   &  2454326.8701  & 0.0013   &           170.7 & 5.1  & $-$23.0 & 4.0  & 27.3 \\
\hline \end{tabular} \end{center} \end{table*}

\subsection{Radial velocity measurement}                                                  \label{sec:data:rv}

We measured radial velocities (RVs) from emission lines in the spectra of our targets\footnote{The reduced spectra and RVs presented in this work will be available at the CDS ({\tt http://cdsweb.u-strasbg.fr/}) and at {\tt http://www.astro.keele.ac.uk/$\sim$jkt/}} by cross-correlation with single and double Gaussian functions \citep{SchneiderYoung80apj,Shafter83apj}, as implemented in {\sc molly}. In each case we tried a range of different widths and separations for the Gaussians in order to verify the consistency of our results (see Paper\,I for further details).

\subsection{Orbital period measurement}                                             \label{sec:data:period}

The RVs and light curves for each CV were searched for periods using periodograms computed by the \citet{Scargle82apj} method, analysis of variance \citep[AoV; ][]{Schwarzenberg89mn} and orthogonal polynomials \citep[ORT; ][]{Schwarzenberg96apj}, as implemented within the {\sc tsa}\footnote{\tt http://www.eso.org/projects/esomidas/doc/user/98NOV\\/volb/node220.html} context in {\sc midas}. In general two Fourier terms were used for ORT, which is appropriate for the relatively simple variation exhibited by these objects.

To find the final values of the orbital periods, and to investigate the aliases in the periodograms, we fitted circular spectroscopic orbits (sine curves) to the data using the {\sc sbop}\footnote{Spectroscopic Binary Orbit Program, written by P.\ B.\ Etzel, \\ {\tt http://mintaka.sdsu.edu/faculty/etzel/}} program, which we have previously found to give reliable error estimates \citep{Me+05mn}. The parameters of the final spectroscopic orbits are given in Table\,\ref{tab:orbits}, and greyscale plots of the trailed spectra are shown in Fig.\,\ref{fig:vlt:trailed}.

In some cases, to assess the likelihood of the point of highest power corresponding to the actual orbital period we have performed bootstrapping simulations (see Paper\,I) by randomly resampling the data with replacement and calculating a new periodogram a total of 1000 times \citep[][p.\,686]{Press+92book}. The fraction of periodograms in which the highest peak fell close to a particular alias can be interpreted as the likelihood of that alias being correct. However, in these cases we expect the resulting probabilities for the correct peak to be quite conservative (i.e.\ too low) for two reasons. Firstly, because the simulated datasets necessarily contain fewer unique epochs than the original data some temporal definition is sacrificed. The bootstrapping periodograms are often clearly inferior to those calculated from the actual data, confirming this picture. Secondly, when picking the best alias interactively we use more information than just the periodogram (including the shape of the phased radial RV curve).

\reff{At the suggestion of the referee we have also performed Monte Carlo simulations, using the same approach as in \citet{Me++04mn2,Me+04mn3}. These have the advantage that they do not suffer from a loss of time information, but the drawback that a sine curve is assumed to be a good representation of the data. This is not always the case for CVs, which often has substantial non-Keplerian kinematical effects. We find, as expected, that Monte Carlo simulations generally give a higher significance level (by a factor of two in precentage terms) to the highest peak in a periodogram.}

\section{RESULTS FOR EACH SYSTEM}

\subsection{SDSS J004335.14$-$003729.8}                                              \label{sec:0043}

SDSS\,J0043 was found to be a CV by \citet{Szkody+04aj} from an SDSS spectrum which shows a blue continuum with narrow Balmer emission lines. There are very broad Balmer absorption lines attributable to the white dwarf component of the system, but no identifiable features arising from the secondary star. \citet{Szkody+04aj} obtained nine low-resolution spectra of SDSS\,J0043 over 2.7\,h using the 3.5\,m telescope and Double Beam Spectrograph at Apache Point Observatory (APO). RVs from the H$\alpha$ and H$\beta$ emission lines yielded orbital periods of about 1.5\,h and 1.2\,h, respectively.

\begin{figure} \includegraphics[width=0.48\textwidth,angle=0]{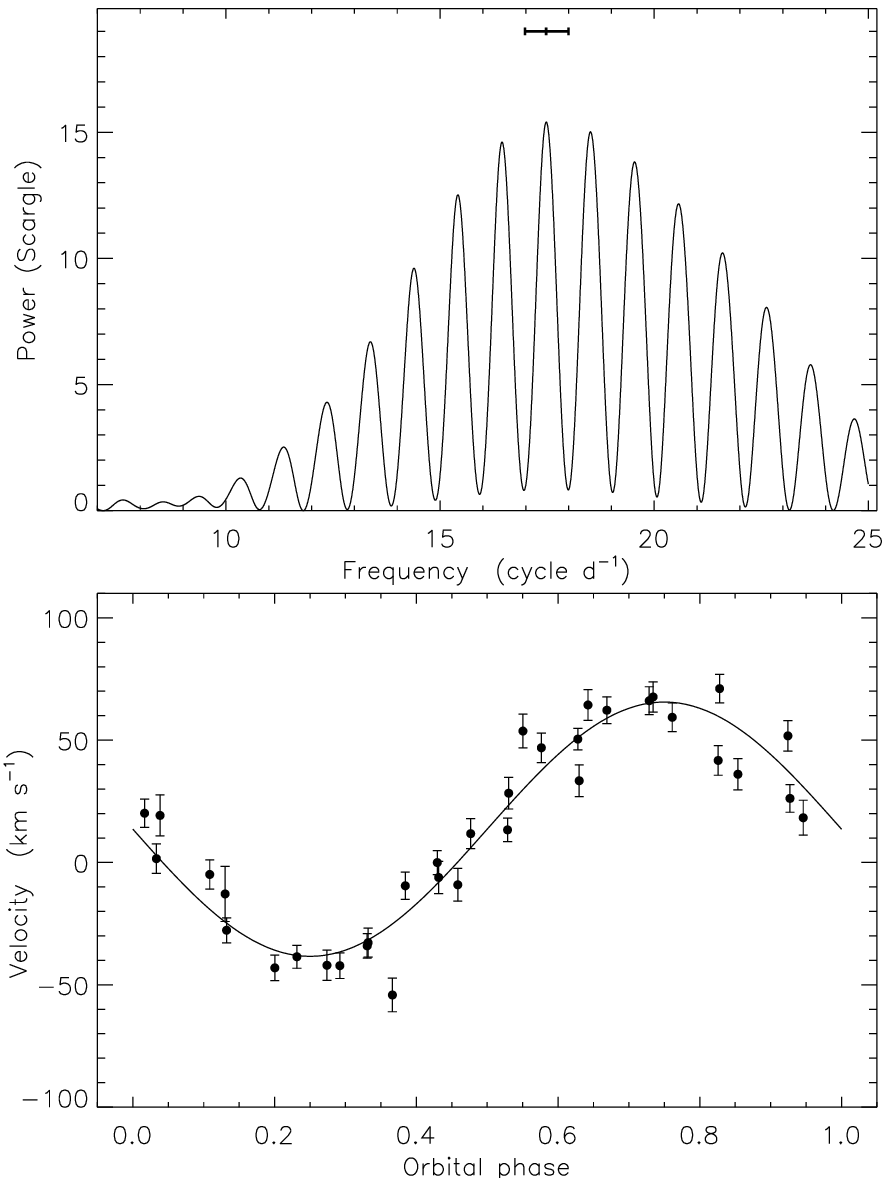}\\
\caption{\label{fig:0043:rvplot} {\it Upper panel:} Scargle periodogram of
the RVs of SDSS\,J0043 measured using a single Gaussian with width 1500\kms.
The measured period and uncertainty from data from the first night only are
indicated with an errorbar at the top of the plot. {\it Lower panel:}
measured RVs (filled circles) compared to the best-fitting spectroscopic
orbit (solid line).} \end{figure}

We obtained a total of 35 VLT spectra of SDSS\,J0043 (21 on one night and 14 on the next night). The H$\alpha$ line is relatively narrow (about 1000\kms) but is clearly double-peaked and variable (Fig.\,\ref{fig:Halpha}). RV measurements consistently give orbital periods close to 82\,min, and the best results are found using a single Gaussian of width 1500\kms. Data from the first night alone give $\Porb = 82.4 \pm 2.4$\,min. Including the spectra from the second night too gives $\Porb = 82.325 \pm 0.088$\,min. The adjacent one-day aliases at 77.75 and 87.49 min differ from the first-night period by 2\hspace*{1pt}$\sigma$, so are unlikely to be correct but cannot be ruled out. Bootstrapping simulations (which we expect to be conservative) give a probability of 75\% that we have identified the orbital period correctly, and respectively 6\% and 19\% that the peaks at 77.75 and 87.49 min are actually the orbital period. \reff{Monte Carlo simulations give a probability of 89\% for this period, and probabilities of 2\% and 9\% for the other peaks.} The parameters of the final spectroscopic orbit are given in Table\,\ref{tab:orbits} and the orbit and Scargle periodogram are plotted in Fig.\,\ref{fig:0043:rvplot}.

A diagram of the phase-binned trailed spectra of SDSS\,J0043 shows that the relative strengths of the double peaks is variable (Fig.\,\ref{fig:vlt:trailed}) and that there is emission in the form of an S-wave from the bright spot on the edge of the accretion disc \citep{Smak85aca}. The \ion{He}{I} 6678\,\AA\ emission line is narrow and single-peaked (Fig.\,\ref{fig:vlt:trailed}) and has the same phasing as the emission from the bright spot in H$\alpha$.

\subsubsection{Photometry of SDSS\,J0043: does it contain a pulsating white dwarf?}       \label{sec:0043:phot}

\begin{figure} \includegraphics[width=0.48\textwidth,angle=0]{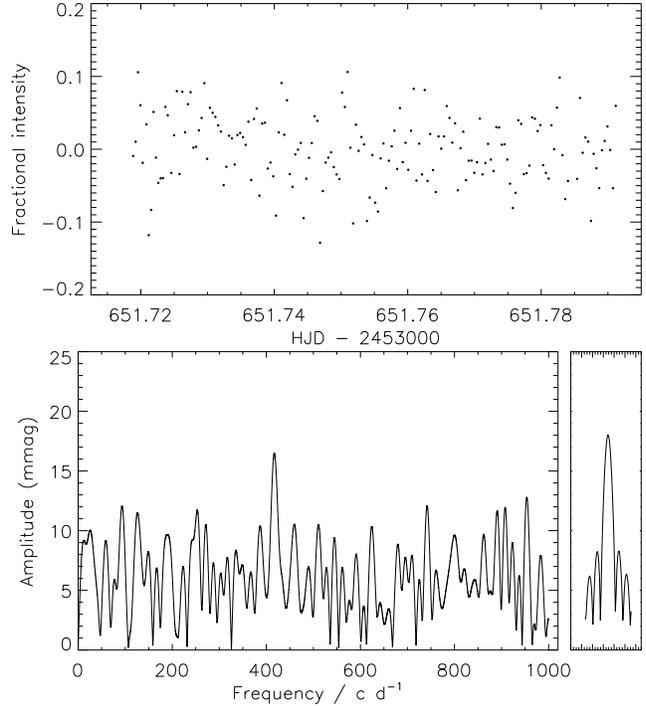}\\
\caption{\label{fig:0043:lcplot} {\it Upper panel:} Light curve of SDSS\,J0043
given as fractional change in intensity. {\it Lower panels:} Amplitude spectrum
of the light curve (left) and the window function (right).} \end{figure}

Fig.\,\ref{fig:0043:lcplot} shows the light curve we have obtained of SDSS\,J0043. A periodogram of these data shows a peak at a frequency near to 400\cd. A fit to this peak results in a period of $207 \pm 1$\,s and an amplitude of $17 \pm 5$\,mmag. This period is within the range of values typically exhibited by ZZ\,Ceti stars (pulsating white dwarfs), but the amplitude is large for this type of star \citep{Winget98jpcm,Mukadam+04apj2}. These observations therefore suggest that the white dwarf in SDSS\,J0043 is a ZZ\,Ceti star. More extensive data are needed for confirmation.

\subsubsection{Doppler tomography of SDSS\,J0043}                                         \label{sec:0043:map}

\begin{figure} \includegraphics[width=0.48\textwidth,angle=0]{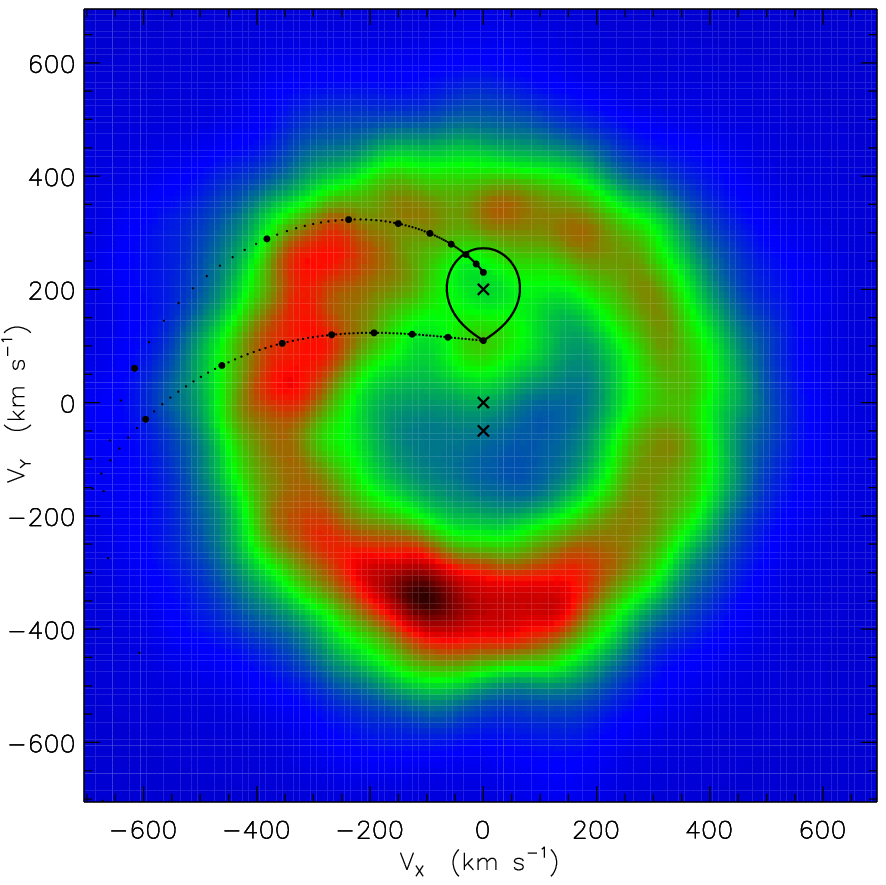}\\
\caption{\label{fig:0043:map} Doppler map of the H$\alpha$ emission line from
SDSS\,J0043. Strong emission is coloured red and weak to no emission is blue. The
Roche lobe of the mass donor is indicated by a solid curve, and the centre of mass
of the system and individual stars are shown with crosses. The dots indicate the
velocity of the accretion stream and the Keplerian velocity of the disc along the
path of the stream. They are positioned at every $0.01R_{\rm RL}$, decreasing from
$R_{\rm RL} = 1.0$ at the mass donor, where $R_{\rm RL}$ is the radius of the Roche
lobe of the white dwarf.} \end{figure}

\begin{figure} \includegraphics[width=0.48\textwidth,angle=0]{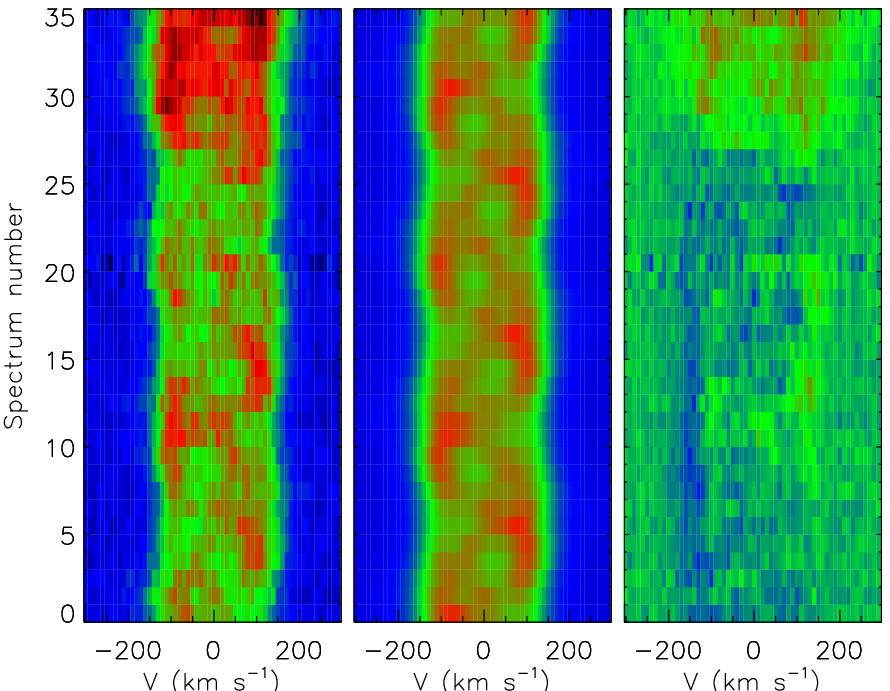}\\
\caption{\label{fig:0043:mapfit} Comparison between the 35 observed spectra
(left panel) and the best-fitting representation for the Doppler map (centre).
The residuals (right) are shown in the right-hand panel.} \end{figure}

The complexity of the H$\alpha$ emission line observed for SDSS\,J0043 prompted us to decompose the emission into velocity space using Doppler tomography. A Doppler map of this line was computed using the maximum entropy method \citep{MarshHorne88mn} and is shown in Fig.\,\ref{fig:0043:map}. \reff{The $\chi^2$ value for the Doppler map was chosen to be marginally larger than the value for which noise features start to be visible. SDSS\,J0043 is not eclipsing, so the exact orientation of the map is not known. The velocity modulation of emission lines from the accretion discs of CVs are habitually offset in phase from the true motion of the white dwarf \citep[e.g.][]{Stover81apj,Steeghs+07apj}. We therefore adopted the spectroscopic orbit and experimented with phase offsets of 0.10--0.25, a range which covers the values found in most studies of eclipsing CVs.} The best-fitting calculated spectra and residuals are shown in Fig.\,\ref{fig:0043:mapfit}.

\reff{The Doppler map in Fig.\,\ref{fig:0043:map} shows a circular emission feature at large velocities which comes from the accretion disc of SDSS\,J0043. There is no emission at low velocities which could be attributed to the white dwarf. However, there is an emission peak at $(V_X,V_Y) = (100,0)$ which may arise from the irradiated inner face of the secondary star. We have overplotted several features on the Doppler map to illustrate this interpretation. The Roche lobe of the secondary is shown with a solid line, the centres of mass of the system and of the two stars are shown by crosses, and the velocity of the accretion stream and the Keplerian velocity of the accretion disc are indicated by dots with a constant spacing in position. We have adopted  $K_{\rm WD} = 50$\kms\ for the white dwarf velocity amplitude (Table\,\ref{tab:orbits}). To position the inner Lagrangian point of the secondary on the emission peak we used a phase offset of 0.21 (corresponding to a clockwise rotation of the map of 76$^\circ$) and a secondary star velocity amplitude of $K_2 = 200$\kms.} This interpretation leads to a mass ratio of $q = \frac{M_2}{M_{\rm WD}} = 0.25$, which is larger than expected and unlikely to be real. We discuss this further below.

The accretion disc has two regions of enhanced emission in the Doppler map. The one at $(V_X,V_Y) = (-350,100)$ is in the path of the accretion stream from the secondary star's surface at the inner Lagrangian point, supporting our interpretation of the map. Doppler maps of helium emission can be useful indicators of the velocity of this bright spot, so a map of the \ion{He}{I} 6678\,\AA\ emission was constructed (not shown) using the same procedure as for the H$\alpha$ map. This displays only one significant feature: a spot of emission centred on roughly $(V_X,V_Y) = (-250,150)$ which confirms that this emission region is arising from the bright spot in SDSS\,J0043.

The second region of enhanced emission from the accretion disc, at $(V_X,V_Y) = (-100,-350)$, is not a typical feature of Doppler maps of CV emission lines. This has no immediate explanation in the accepted picture of the structure of short-period CVs, but has previously been seen in some AM\,CVn system \citep[e.g.][]{Roelofs+06mn}. A second oddity for this system is that the placement of the inner emission feature on the substellar point of the secondary star required $K_2 = 200$\kms, which with $K_{\rm WD} = 50$\kms\ results in $q = 0.25$, much larger than the expected value of $q \sim 0.1$ for a CV with $\Porb = 82$\,min \citep{Knigge06mn}. The latter difficulty could be explained by the former: the region of enhanced emission at $(V_X,V_Y) = (-100,350)$ will distort the H$\alpha$ profiles and thus the velocities measured from them. The spectroscopic orbit measured above is therefore unlikely to accurately represent the motion of the white dwarf. This caveat is supported by the phase difference of 0.21 between the spectroscopic orbit we measured from the H$\alpha$ emission and the velocity variation of the irradiated face of the secondary star, which in our experience is on the large side for short-period CVs. If $K_{\rm WD}$ were only 25\kms\ then the mass ratio would become roughly $q = 0.12$, which is a reasonable value for this system. An alternative explanation is that the degenerate component has a low mass of $M_{\rm WD} \sim 0.4$\Msun, which would make it a helium-core white dwarf.


\subsection{SDSS J033710.91$-$065059.4}                                                   \label{sec:0337}

\begin{figure} \includegraphics[width=0.48\textwidth,angle=0]{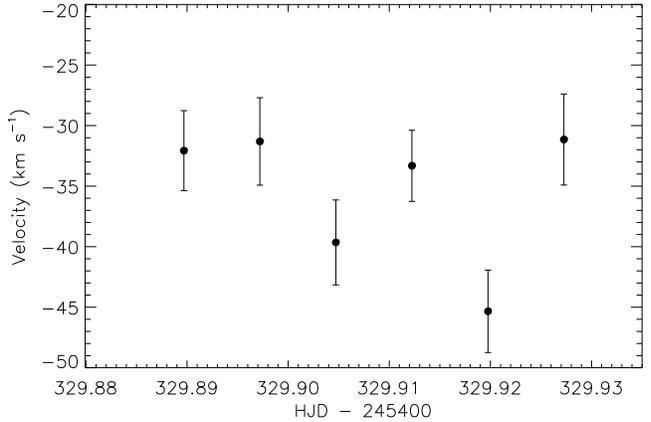}\\
\caption{\label{fig:0337:rvplot} RVs measured for SDSS\,J0337 by
cross-correlation against a single Gaussian function with FWHM 600\kms.}
\end{figure}

A spectrum of SDSS\,J0337 was presented by \citet{Szkody+07aj} which identified it as a CV with narrow Balmer emission lines. The apparent magnitude of SDSS\,J0337 was much brighter when it was observed during the SDSS imaging observations ($g = 19.54$) than when the SDSS spectrum was subsequently acquired ($g_{\rm spec} = 23.3$). This is probably due to the system being in a higher state during the imaging observations, an explanation which is supported by the brightness of SDSS\,J0337 in our acquisition image ($V \approx 21.4$) being midway between the two SDSS magnitudes.

SDSS\,J0337 was the last object studied during our VLT observing run, and there was only time to obtain seven spectra over one hour. These show a sharp H$\alpha$ emission line (Fig.\,\ref{fig:Halpha}) with a FWHM of only about 500\kms, with weaker narrow \ion{He}{I} emission lines at 5876\,\AA, 6678\,\AA\ and 7065\,\AA. We cannot detect any RV motion in the H$\alpha$ line, to an upper limit of 15\kms. The narrow emission lines and lack of observable RV variation are consistent with a low orbital inclination for the SDSS\,J0337 system. A plot of its RVs, measured using a single Gaussian of width 600\kms, is shown in Fig.\,\ref{fig:0337:rvplot}. \reff{Obtaining the orbital period of this object may take a substantial amount of telescope time.}


\subsection{SDSS J160111.53$+$091712.6}                                                   \label{sec:1601}

\begin{figure} \includegraphics[width=0.48\textwidth,angle=0]{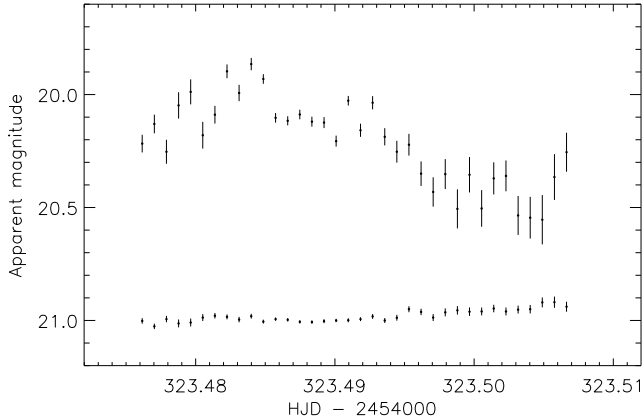}\\
\caption{\label{fig:1601:lcplot} NTT unfiltered photometry of SDSS\,J1601. The
magnitudes are differential with respect to a comparison star and have been
offset by the $r$ magnitude of the comparison star. Differential magnitudes for
the comparison minus check stars are shown offset by +21.3\,mag.} \end{figure}

SDSS\,J1601 was discovered to be a CV by \citet{Szkody+06aj} and has a spectrum characterised by strong Balmer and weak \ion{He}{I} emission lines which are slightly double-peaked (Fig.\,\ref{fig:sdssspec}). We obtained 45\,min of NTT unfiltered photometry, which shows a variation of amplitude about 0.2--0.3\,mag (Fig.\,\ref{fig:1601:lcplot}). The mean magnitude is consistent with the SDSS imaging and spectroscopic values. \reff{The variation is suggestive of a sinusoidal variation with a period similar to the duration of the observations}. Further photometry of SDSS\,J1601 has a good chance of yielding a measurement of the orbital period of this binary.


\subsection{SDSS J163722.21$-$001957.1}                                                   \label{sec:1637}

\begin{figure} \includegraphics[width=0.48\textwidth,angle=0]{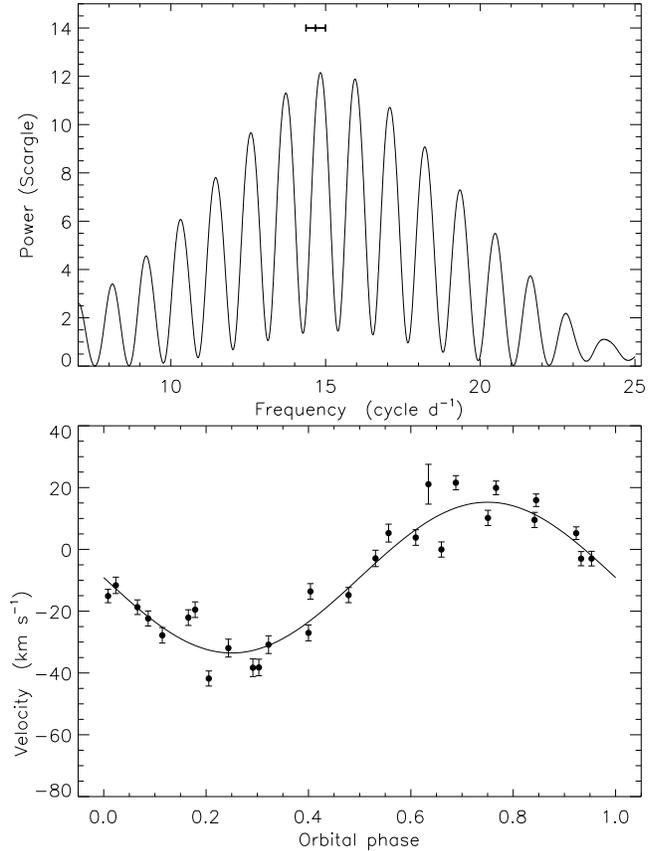}\\
\caption{\label{fig:1637:rvplot} {\it Upper panel:} Scargle periodogram of
the RVs of SDSS\,J1637 measured using a single Gaussian with width 500\kms.
The measured period and uncertainty from data from the first night only are
indicated with a thick line. {\it Lower panel:} measured RVs (filled circles)
compared to the best-fitting spectroscopic orbit (solid line).} \end{figure}

SDSS\,J1637 is a faint CV ($V \approx 20.5$) which was in a high state when observed by the SDSS imaging survey ($g = 16.60$). The flux level of its SDSS spectrum \citep{Szkody+02aj} ($g_{\rm spec} = 20.6$) is similar to its apparent magnitude in our VLT acquisition observations ($V = 20.3$ and 20.6). \citet{Szkody+02aj} observed the system several times (the exact number is not given) and found it at 20th magnitude each time. A superoutburst of SDSS\,J1637 was observed by G.\ Bolt\footnote{{\tt vsnet-superoutburst} alert number 2306} with a mean magnitude of $V \approx 15.5$ on the night of 2004 March 28. A superhump period of $0.06927 \pm 0.0006$\,d ($99.75 \pm 0.86$\,min) was measured by G.\ Bolt from CCD photometry\footnote{{\tt vsnet-superoutburst} alert number 2310}.

We obtained 28 VLT spectra of SDSS\,J1637 over two nights, 20 of which were taken over three hours on the first night. These show a strong H$\alpha$ emission line with FWHM of only about 800\kms, and much weaker \ion{He}{I} emission at 6678\,\AA\ and 7065\,\AA. RV measurements with both single and double Gaussians of various widths are in good agreement with each other, and formally the best results are obtained using a single Gaussian of FWHM 500\kms. RV measurements from the first night yield a period of $98.2 \pm 2.1$\,min. Adding in the second night gives a best period of 97.0\,min with one-day aliases at 90.2 and 104.9 min. Using the measurements from the first night to select the best alias gives an orbital period of $97.04 \pm 0.19$\,min (Fig.\,\ref{fig:1637:rvplot}). Bootstrapping simulations give a (conservative) probability of 80\% that the orbital period refers to the 97-min peak in the periodogram, and a 19\% probability that the 105-min alias is actually the correct period. \reff{Monte Carlo simulations are more confident, yielding an 88\% probability for the 97-min peak and a 10\% probability for the 105-min alias.} The 97-min period receives further support from the observed superhump period for SDSS\,J1637, so can be unambiguously assigned to the orbital period of the system.

\subsubsection{The physical properties of SDSS\,J1637}

\reff{There is strong evidence that} the superhumps observed in CVs during superoutburst arise from the precession of an elliptical accretion disc \citep{Vogt82apj,Whitehurst88mn}, where the superhump period is the beat period between the orbital and precession periods. The ratio of the orbital and superhump periods is observed to depend on mass ratio in CVs \citep{Patterson98pasp}, allowing the physical properties of SDSS\,J1637 to be estimated.

From measurements of the orbital and superhump periods of SDSS\,J1637 we find a superhump period excess \citep{Patterson98pasp} of $\epsilon(q) = 0.028 \pm 0.009$. Using the calibration presented by \citet{Patterson98pasp} we obtain from this $q = 0.167$. The updated calibration given by \citet{Knigge06mn} results in $q = 0.161 \pm 0.037$, which is in agreement (but remember that the different calibrations have many objects in common). \reff{Estimating a} white dwarf mass of $M_{\rm WD} = 0.8$\Msun\ \citep{SmithDhillon98mn,Littlefair+08} results in a secondary star mass of $M_2 = 0.13$\Msun, in agreement with the properties of the semi-empirical CV secondary star sequence presented by \citet{Knigge06mn}.

The measured spectroscopic orbit for SDSS\,J1637 has a low velocity amplitude of only $24.4 \pm 1.6$\kms\ (Table\,\ref{tab:orbits}). The phase-binned and trailed H$\alpha$ and \ion{He}{I} 6678\,\AA\ spectra show no large variation with orbital phase (Fig.\,\ref{fig:vlt:trailed}). SDSS\,J1637 has the largest H$\alpha$ and \ion{He}{I} emission line equivalent widths of all of the objects we have studied with the VLT, which points to a very weak continuum. The H$\alpha$ emission line is strong, narrow and single-peaked. These results suggest that the SDSS\,J1637 binary system has a relatively low orbital inclination when observed from Earth. If the masses of the white dwarf and secondary star are 0.8\Msun\ and 0.13\Msun\ (see above), the velocity amplitude we have measured implies an orbital inclination of roughly 40$^\circ$.




\subsection{SDSS J164248.52$+$134751.4}                                                   \label{sec:1642}

\begin{figure} \includegraphics[width=0.48\textwidth,angle=0]{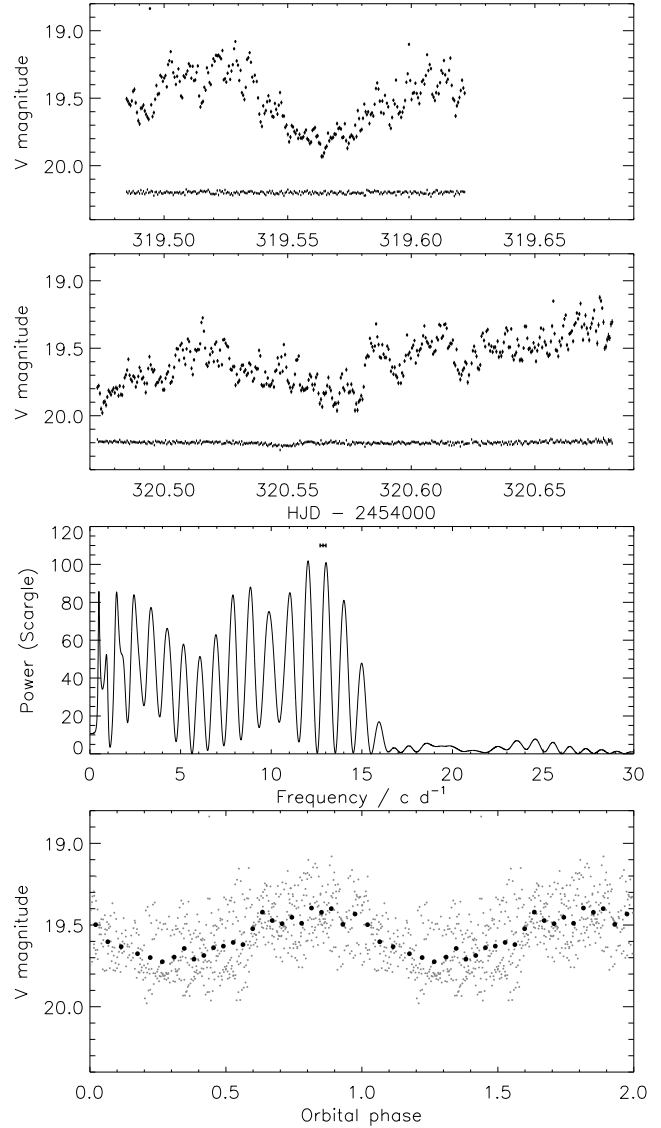}\\
\caption{\label{fig:1642:lcplot} NTT $V$-filter photometry of SDSS\,J1642. The upper two
panels show the light curves from the individual nights as well as differential photometry
between the two comparison stars used (offset to appear in the plot). The third panel shows
the Scargle periodogram of the combined data from the two nights, and the spectroscopic
orbital period value is indicated at the top of the plot. The bottom panel shows the data
phased with the photometric period of 110.6\,min (grey dots) and combined into 25 phase bins
(black filled circles). In most cases the error bars are smaller than the points.} \end{figure}

\begin{figure} \includegraphics[width=0.48\textwidth,angle=0]{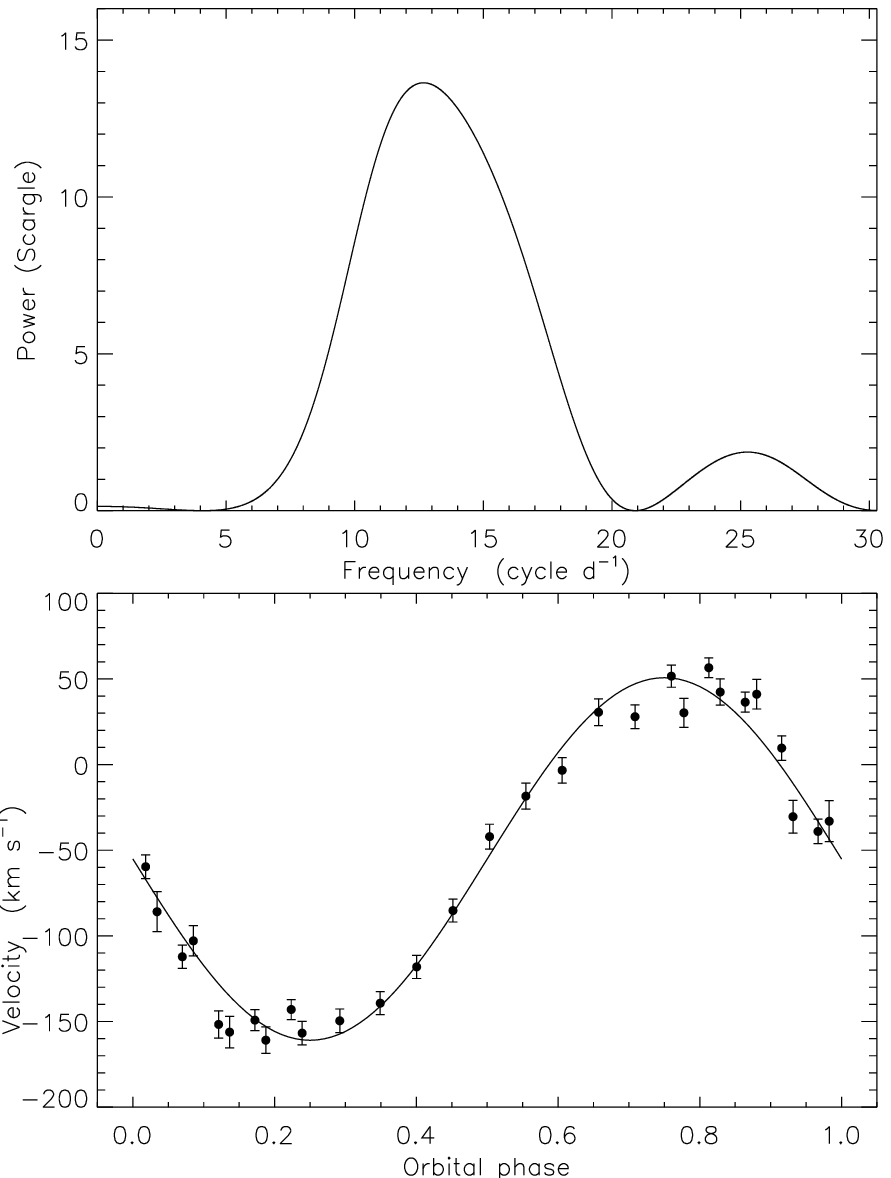}\\
\caption{\label{fig:1642:rvplot} {\it Upper panel:} Scargle periodogram of
the RV of SDSS\,J1642 measured using a double Gaussian with width 300\kms\
and separation 2100\kms. {\it Lower panel:} measured RVs (filled circles)
compared to the best-fitting spectroscopic orbit (solid line).} \end{figure}

SDSS\,J1642 was discovered to be a CV by \citet{Szkody+07aj} and has double-peaked Balmer and \ion{He}{I} emission lines. Weak emission at \ion{He}{ii} 4686\,\AA\ is also visible in the SDSS spectrum. \citet{Szkody+07aj} presented seven low-resolution APO 3.5\,m telescope spectra taken over an 80\,min period and \reff{estimated} an orbital period ($70 \pm 7$\,min, which is close to the duration of the observations) from RV measurements of the H$\alpha$ and H$\beta$ lines.

We obtained eight hours of $V$-filter photometry of SDSS\,J1642 over two nights in 2007 August using the NTT and SUSI2 imager. Exposure times of 30\,s were used, giving an observing cadence of 43\,s. This object is quite variable in brightness: it was at magnitude $g = 18.64$ in the SDSS imaging data, $g_{\rm spec} = 18.0$ during its SDSS spectrum, averaging $V = 19.5$ on the night of 2007 August 06, and during the next night increased from $V = 19.8$ to $V = 19.3$ over five hours. Its apparent magnitude on our VLT acquisition image on the night of 2007 August 17 was $V = 18.5$.

The light curves from the two NTT nights are shown in Fig.\,\ref{fig:1642:lcplot}. Periodic variation is seen during both nights, as well as a steady increase in brightness over the second night and quite a lot of flickering \citep{Bruch92aa,Bruch00aa}. Scargle periodograms show a small forest of peaks in the region between zero and 15 cycles per day (Fig.\,\ref{fig:1642:lcplot}, panel 3) and it is not clear for these data if any of the peaks relates to the orbital period.

We observed SDSS\,J1642 for three hours with VLT/FORS2 on the night of 2007 August 17, obtaining 29 spectra with exposure times of 300\,s. The H$\alpha$ emission line is strong and has double peaks separated by 750\kms\ (measured from the average spectrum). The \ion{He}{I} 6678\,\AA\ line is weaker and its double peaks have a greater separation of 1220\kms. The best RV measurements were obtained using the double Gaussian method with FWHMs 300\kms\ and separation 2100\kms, giving an orbital period of $\Porb = 113.6 \pm 1.5$\,min (Table\,\ref{tab:orbits}).

The RV curve is not sinusoidal (Fig.\,\ref{fig:1642:rvplot}) and formally the best orbital fit has an eccentricity of $e = 0.09 \pm 0.03$ (giving $\sigma_{\rm rms} = 11.2$ compared to $\sigma_{\rm rms} = 12.7$ for a circular orbit) and an orbital period of 113.4\,min. It is difficult to find a reason for an interacting close binary star to have an eccentric orbit, and we do not believe that this is the case for SDSS\,J1642. Many CVs have emission-line RV variations which do not match the motion of the white dwarf component \citep{Stover81apj,Steeghs+07apj}, and an apparently significant nonzero orbital eccentricity may result from the same distortion mechanisms. We therefore adopt the orbital period from the circular RV orbit, which in this case is in excellent agreement with the period from the eccentric-orbit alternative. The spectra have been phase-binned with this period and shown as a trailed greyscale plot in Fig.\,\ref{fig:vlt:trailed}. The emission shows clear double peaks at both H$\alpha$ and \ion{He}{I} 6678\,\AA.

Using the orbital period we measured from the VLT spectra we can now select the peak(s) in the light curve periodogram which may correspond to this value. The two highest peaks are at periods of 110.6 and 119.7 min (Fig.\,\ref{fig:1642:lcplot}), which agree with the spectroscopic period to within 2\,$\sigma$ and 4\,$\sigma$, respectively. The former of these has a period of $110.60 \pm 0.16$\,min. This might represent the orbital period of the system, but the data in hand are insufficient to be sure. We therefore adopt the spectroscopic value, $\Porb = 113.6 \pm 1.5$\,min, as the orbital period of SDSS\,J1642.

\subsubsection{Doppler tomography of SDSS\,J1642}

\begin{figure*}
\includegraphics[width=0.32\textwidth,angle=0]{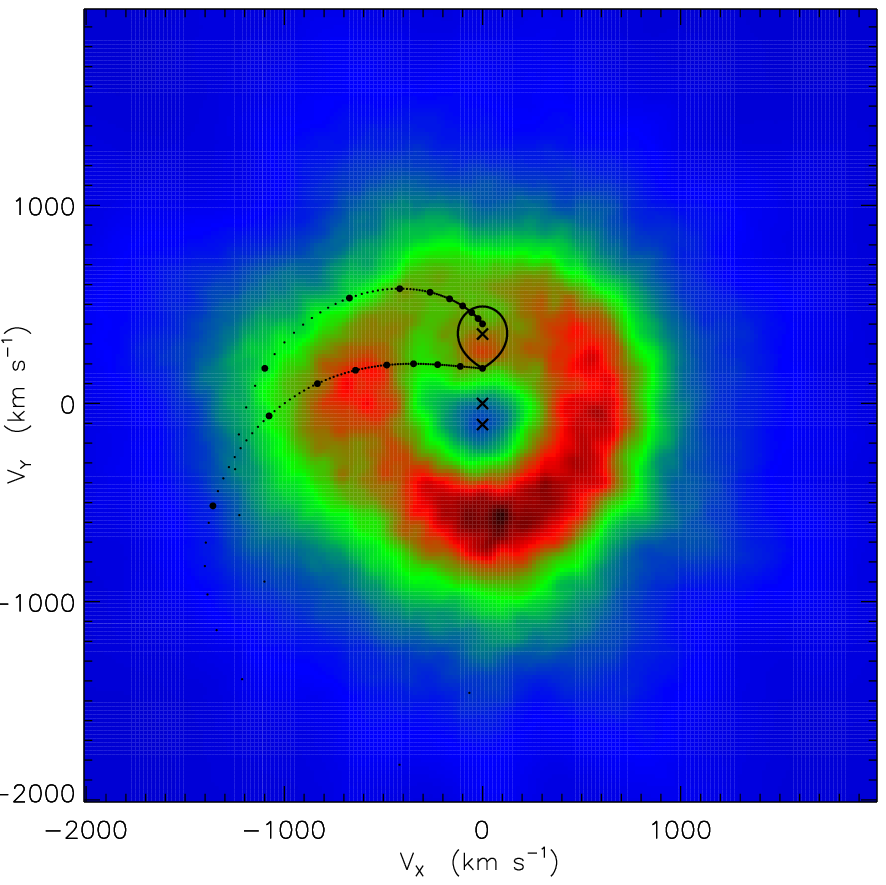} \ \
\includegraphics[width=0.32\textwidth,angle=0]{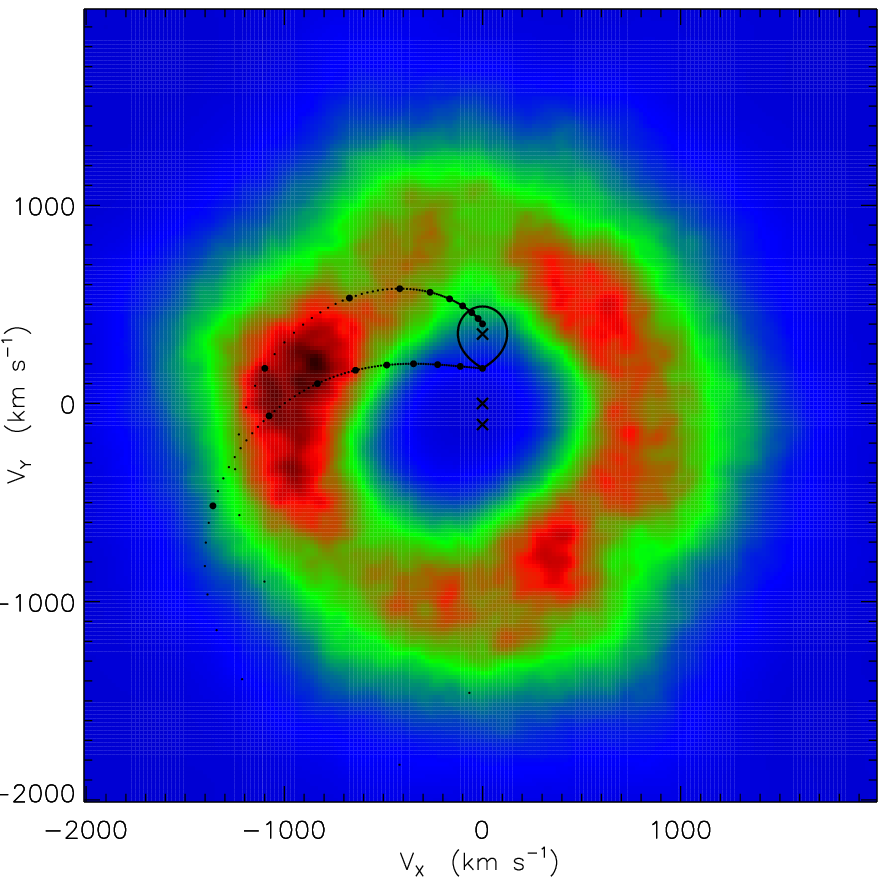} \ \
\includegraphics[width=0.32\textwidth,angle=0]{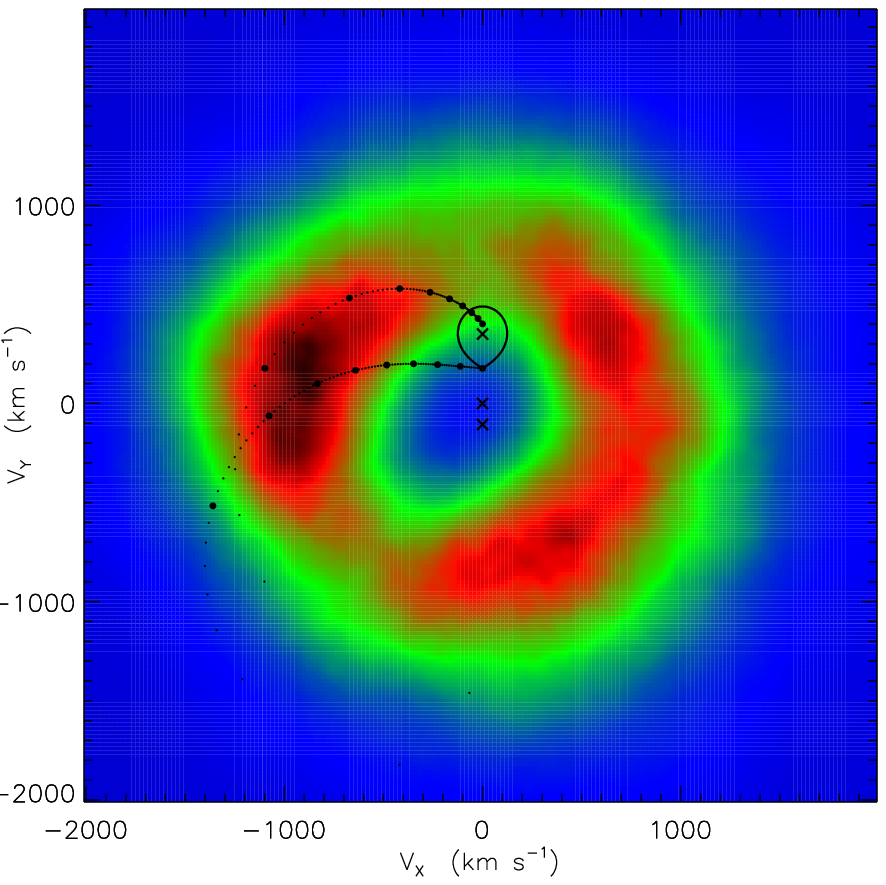}\\
\caption{\label{fig:1642:map} Doppler maps of the H$\alpha$ (left),
\ion{He}{I} 6678\,\AA\ (centre) and \ion{He}{I} 7065\,\AA\ (right)
emission lines from SDSS\,J1642. The symbols are overplotted in the
same way as for Fig.\,\ref{fig:0043:map}.} \end{figure*}

The trailed spectra for SDSS\,J1642 (Fig.\,\ref{fig:vlt:trailed}) show that the \ion{He}{I} 6678\,\AA\ emission line appears to be wider than the H$\alpha$ line. We have therefore calculated Doppler maps for SDSS\,J1642, in the same way as for SDSS\,J0043 (Section\,\ref{sec:0043:map}), to investigate this further. The maps are plotted in Fig.\,\ref{fig:1642:map} and indeed show that the accretion disc seems to span a wider range of velocities in \ion{He}{I} 6678\,\AA\ and \ion{He}{I} 7065\,\AA\ than for H$\alpha$. These higher velocities imply that the helium emission comes from a physically smaller part of the disc, specifically the hotter annuli closer to the white dwarf surface.

A second feature of the H$\alpha$ Doppler map (but not the \ion{He}{I} maps) is an inner emission feature that, \reff{by analogy with SDSS\,J0043 and many other CVs}, can be attributed to the surface of the secondary star. The interpretations of the maps in  Fig.\,\ref{fig:1642:map} assume $K_2 = 350$\kms\ and a phase offset of 0.15. In this interpretation, the lower velocities of the accretion disc in the H$\alpha$ map overlap that of the inner emission feature, which violates Kepler's third law. \reff{The disparity between the H$\alpha$ and \ion{He}{I} Dopper maps, and the apparent violation of Kepler's third law, can both be explained by the breakdown of the assumption that all emitting material is optically thin. If the accretion disc is optically thin to \ion{He}{I} emission but partially optically thick for H$\alpha$, it is quite possible that the detected H$\alpha$ emission is biased towards lower velocities.}

The H$\alpha$ map also shows a region of strong emission coming from the accretion disc, on the side opposite to the bright spot and reminiscent of the Doppler maps of SDSS\,J0043 (Section\,\ref{sec:0043}). Our spectroscopic observations only cover 1.5 orbital periods of this system, and the region of strong emission could be caused by brightness variations which change between different orbits (which is not taken into account when calculating the Doppler maps). Additional observations are needed to further investigate the unusual characteristics of SDSS\,J1642.


\subsection{SDSS J165837.70$+$184727.4}                                                   \label{sec:1658}

\begin{figure} \includegraphics[width=0.48\textwidth,angle=0]{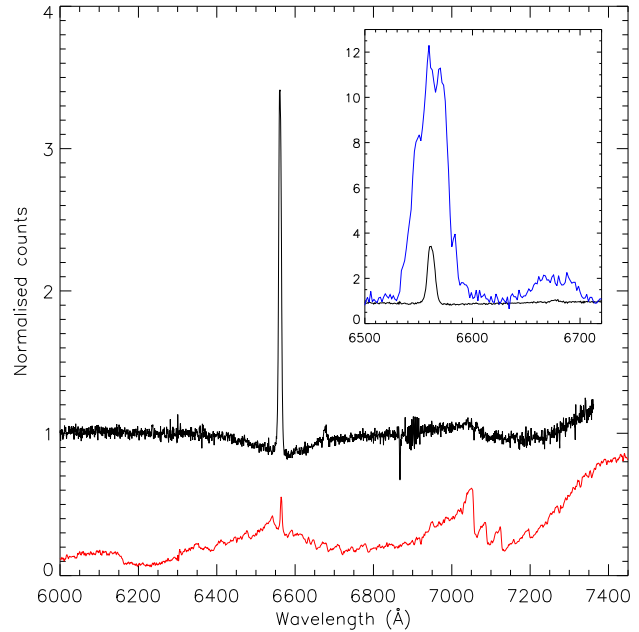}\\
\caption{\label{fig:1658:spec} {\it Main panel:} The mean spectrum of SDSS\,J1658
from our VLT observations (black line), normalised unity using a straight-line fit.
A template M5 star is shown, with a red line, for comparison. {\it Inset panel:}
A comparison between our mean VLT spectrum (black line) and the SDSS spectrum of
SDSS\,J1658 (blue line) in the region of the H$\alpha$ and \ion{He}{I} 6678\,\AA\
lines.} \end{figure}

\begin{figure} \includegraphics[width=0.48\textwidth,angle=0]{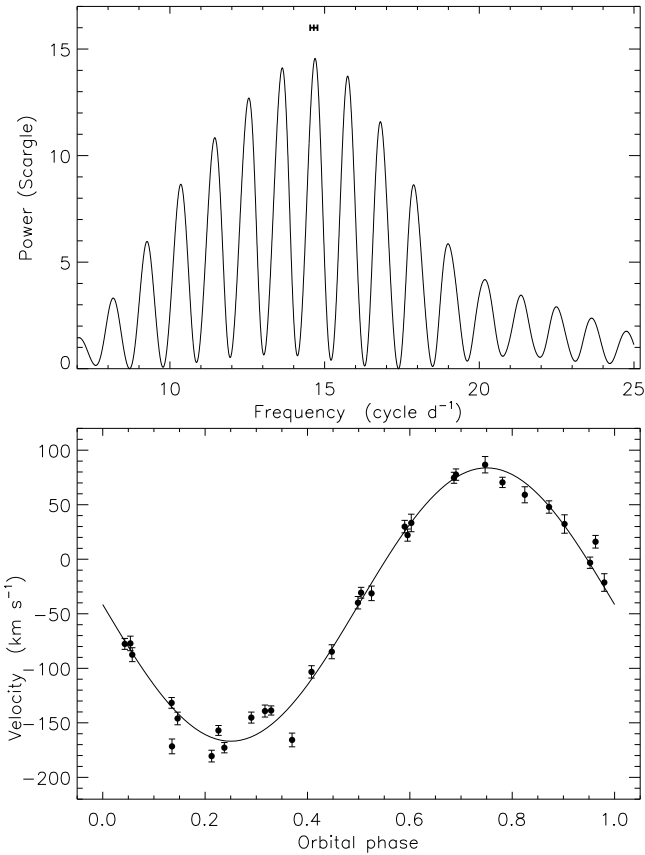}\\
\caption{\label{fig:1658:rvplot} {\it Upper panel:} Scargle periodogram of
the RVs of SDSS\,J1658 measured using a single Gaussian with width 400\kms.
{\it Lower panel:} measured RVs (filled circles) compared to the best-fitting
spectroscopic orbit (solid line).} \end{figure}

SDSS\,J1658 was discovered to be a CV by \citet{Szkody+06aj} and its SDSS spectrum shows the strong Balmer emission and weaker \ion{He}{I} emission lines typical of short-period CVs. The Balmer emission is single-peaked and the \ion{He}{I} emission is weakly double-peaked.

We obtained 31 VLT spectra of SDSS\,J1658 over two consecutive nights in 2007 August. At the times of the SDSS imaging observations and our spectroscopy, the system had an apparent $g$ or $V$ magnitude of $\sim$20. However, the spectroscopic characteristics of SDSS\,J1658 are very different in our observations compared to the SDSS spectrum (where it was at magnitude $g_{\rm spec} = 19.7$). Fig.\,\ref{fig:1658:spec} shows that the emission lines in the SDSS spectrum are remarkably strong and broad (for example, H$\alpha$ has FWHM 35\,\AA\ and equivalent width 360\,\AA) but in the VLT data are extremely weak (7\,\AA\ and 25\,\AA, respectively). Spectral features from both stellar components are visible in the mean VLT spectrum: we see a broad H$\alpha$ absorption from the white dwarf photosphere and a wide 7150--7350\,\AA\ dip which betrays the presence of an M-type secondary star (Fig.\,\ref{fig:1658:spec}).

At the time of the SDSS spectrum SDSS\,J1658 was clearly in a state of much higher accretion level than during the other observations. This is manifested in the vastly stronger emission lines from the accretion disc, but the disc still contributed very little continuum light so the CV was only slightly brighter at that time. The stellar components are not clearly visible in the SDSS spectrum due to the wider emission lines and much lower S/N than our mean VLT spectrum.

The RV motion of the H$\alpha$ line in our spectra is straightforward to detect. Single-Gaussian measurements were made and the best FWHM was found to be 400\kms, giving an orbital period of $98.29 \pm 0.79$\,min from 19 spectra on the first night only. Including the spectra from the second night gives the expected alias pattern centred on a period of 98.05\,min with one-day aliases at 91.48 and 105.70 min. We can therefore unambiguously identify the central peak in the periodogram with the orbital motion of the binary, resulting in a final period of $\Porb = 98.012 \pm 0.065$\,min. The spectroscopic orbit is plotted in Fig.\,\ref{fig:1658:rvplot} and its parameters are given in Table\,\ref{tab:orbits}.

The velocity amplitude we have measured from the H$\alpha$ emission, $K_1 = 125.3 \pm 3.5$\kms, is too high to be attributed to the white dwarf, and is far too narrow to arise from the full accretion disc. We therefore tentatively assign it to the secondary star. \reff{An additional constraint on the system is that the full width at zero intensity of the H$\alpha$ line in the SDSS spectrum (roughly 3200\kms) arises from material in orbit around the white dwarf. We have constructed a constraints diagram in a similar fashion as for SDSS\,J121607.03$+$052013.9 \citep{Me+06mn} and SDSS\,J013132.39$-$090122.2 \citep{Me+07mn2}, and required that our measured velocity amplitude is attributable to some part of the secondary star. We find that there is a small region of parameter space where the system could satisfy all our constraints, involving a large white dwarf mass (1.0\Msun\ or more) and a low orbital inclination (10--30\degr). This solution also allows the secondary star to have a mass near to 0.15\Msun, which is the value expected for a CV with a period of 98\,min \citep{Knigge06mn}. This is only a rough investigation: more detailed studies would require a more accurate measurement of the emission-line full width at zero intensity when the system is again in the high state. An alternative explanation is that the narrow emission comes from some other structure in the system.}

Our VLT data for SDSS\,J1658 have characteristics reminiscent of the spectrum of the low-inclination CV RE\,J1255+266 presented by \citet{Watson+96mn}, which shows wide absorption from the white dwarf and very narrow Balmer emission. In the case of RE\,J1255+266, the narrow emission arises from the accretion disc and has a much lower velocity amplitude than the one we measure for SDSS\,J1658.

\subsection{SDSS J165951.68$+$192745.6}                                                   \label{sec:1659}

\begin{figure} \includegraphics[width=0.48\textwidth,angle=0]{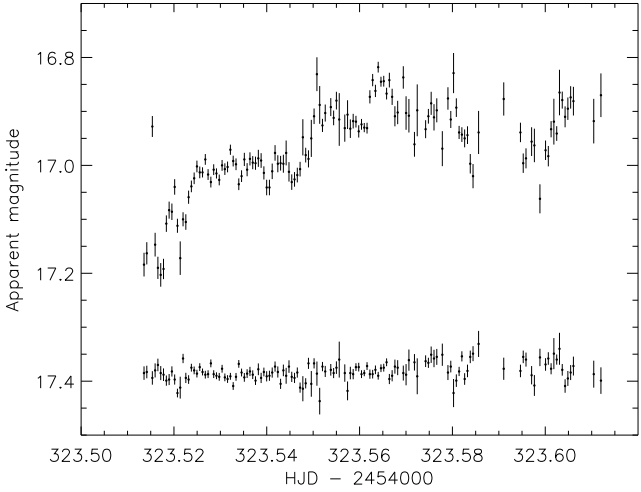}\\
\caption{\label{fig:1659:lcplot} NTT $V$-filter photometry of SDSS\,J1659 obtained
in cloudy conditions. Differential magnitudes for SDSS\,J1659 minus comparison star
are shown as filled circles with error bars, offset by the $V$ magnitude of the
comparison. Differential magnitudes for the comparison minus the check star are
shown offset at the bottom of the plot.} \end{figure}

SDSS\,J1659 was found to be a CV by \citet{Szkody+06aj} from an SDSS spectrum which has the extremely blue continuum and weak single-peaked emission lines which are characteristic of a high mass transfer rate (e.g.\ \citealt{Rodriguez+07mn} and \citealt{Rodriguez+07mn2}). We obtained 2.5\,hours of $V$-filter NTT photometry of SDSS\,J1659 (Fig.\,\ref{fig:1659:lcplot}) as a brighter ($g = 17.12$) backup target during cloudy conditions. There is flickering \citep{Bruch92aa,Bruch00aa} but no clear coherent periodicity in these data.


\subsection{SDSS J223252.35$+$140353.0}                                                   \label{sec:2232}

\begin{figure} \includegraphics[width=0.48\textwidth,angle=0]{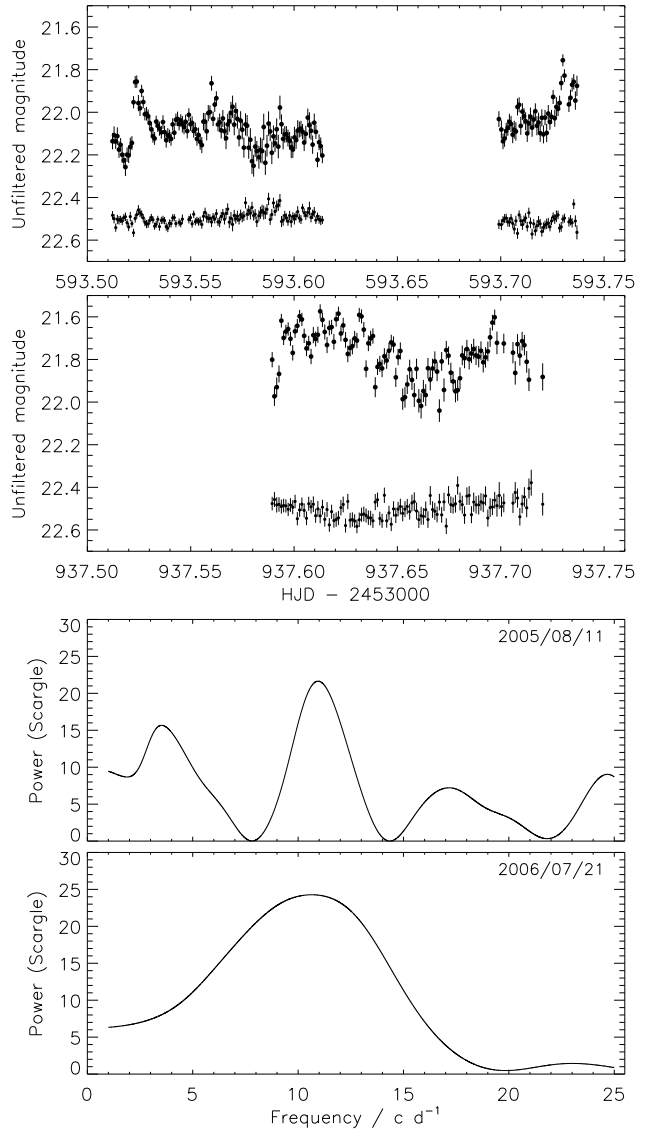}\\
\caption{\label{fig:2232:lcplot} {\it Upper panels:} \reff{NOT unfiltered light curves
of SDSS\,J2232. The comparison minus check magnitudes are shown in each panel
offset to a magnitude of 20.4.} {\it Lower panels:} Scargle periodograms of the
two light curves.} \end{figure}

\begin{figure} \includegraphics[width=0.48\textwidth,angle=0]{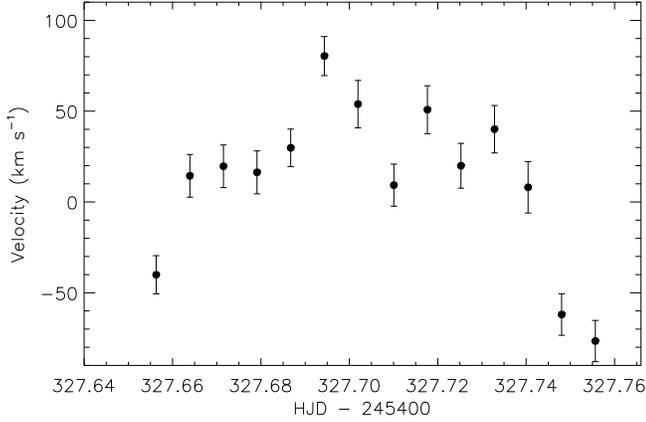}\\
\caption{\label{fig:2232:rvplot} RVs measured for SDSS\,J2232 using a double
Gaussian function with FWHMs 500\kms\ and separation 2000\kms.} \end{figure}

\begin{figure} \includegraphics[width=0.48\textwidth,angle=0]{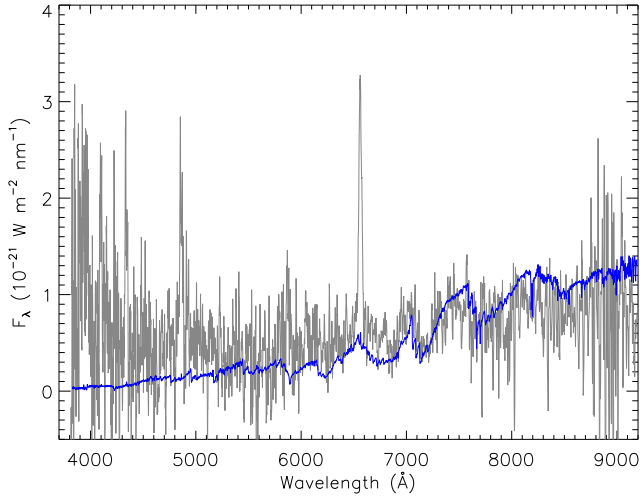}\\
\caption{\label{fig:2232:spec} Compison between the SDSS spectrum of SDSS\,J2232
(grey line; the data have been smoothed) and a template SDSS spectrum of an M4
dwarf (blue line).} \end{figure}

SDSS\,J2232 was found at magnitude $g = 17.7$ by the SDSS imaging survey. It was subsequently selected for spectroscopic follow-up, and the SDSS spectrum \citep{Szkody+04aj} shows Balmer emission lines emanating from a much fainter object ($g_{\rm spec} = 23.2$). This indicates that SDSS\,J2232 is a dwarf nova which was in outburst at the time of the SDSS imaging observations but in quiescence during the SDSS spectroscopic observation. On our VLT acquisition image SDSS\,J2232 was in quiescence at a magnitude of $V = 21.4$, whereas it was at magnitudes 22.1 and 21.8 during our visits to it with the NOT.

The NOT light curves cover time intervals of about three and five hours, separated by slightly less than one year. Both show a variation with a period in the region of 135\,min. The light curves and Scargle periodograms are shown in Fig.\,\ref{fig:2232:lcplot}. Fitting sine curves to the data results in periods of $131 \pm 3$\,min for the 2005 data and \reff{$132 \pm 6$\,min for the 2006 observations}. With the VLT we obtained 14 H$\alpha$ spectra over 2.5 hours, but then discontinued our observations for scheduling reasons. The resulting RVs (Fig.\,\ref{fig:2232:rvplot}) are consistent with orbital motion with a period longer than 4 hours.

The SDSS spectrum of SDSS\,J2232 has a very low signal but has a significant contribution from the secondary star. A simple fit to the the SDSS spectrum (see \citealt{Me+06mn}) indicates a secondary component with a spectral type of M4 and a magnitude $r = 23.4$ \reff{(Fig.\,\ref{fig:2232:spec})}. We would therefore expect to see an ellipsiodal modulation in the light curve of this system, with a period of half of the orbital period. The observed periodicity of roughly 135\,min implies an orbital period of about 270\,min, consistent with the RVs measured from our VLT spectra. We therefore suggest that SDSS\,J2232 is a dwarf nova with $\Porb \sim 4.5$\,hr. These properties are similar to those of the well-studied dwarf novae U\,Gem ($\Porb = 254.7$\,min and secondary spectral type $\sim$M4; \citealt{Naylor++05mn,Echevarria++07aj}) and GY\,Cnc (252.6\,min and M3; \citealt{Gansicke+00aa,Thorstensen00pasp}).

The distance to this system can be estimated 
using Roche geometry. Assuming $M_{\rm WD} = 0.6$ and $q = 0.5$ gives a secondary star radius of about $2.5 \times 10^8$\,m. Accounting for the flux ratio between the SDSS spectrum of SDSS\,J2232 and a template M4 dwarf spectrum (and the flux from the white dwarf and the accretion disc), we find a distance of about 2.7\,kpc. Alternatively, adopting a secondary spectral type of M4 gives an absolute visual magnitude of $M_V = 11.3$. The SDSS magnitude $r = 23.4$ converts to $V = 24.0$, which then results in a distance of roughly 3.5\,kpc. A distance of approximately 3\,kpc puts SDSS\,J2232 about 2\,kpc from the Galactic plane. This is well into the Galactic thick disc, which has a scale height close to 1\,kpc \citep{Veltz+08aa}. Further observations to assess the membership of SDSS\,J2232 in this old stellar population may be very rewarding.



\subsection{SDSS\,J223843.84$+$010820.7 (Aqr\,1)}                                          \label{sec:2238}

\begin{figure} \includegraphics[width=0.48\textwidth,angle=0]{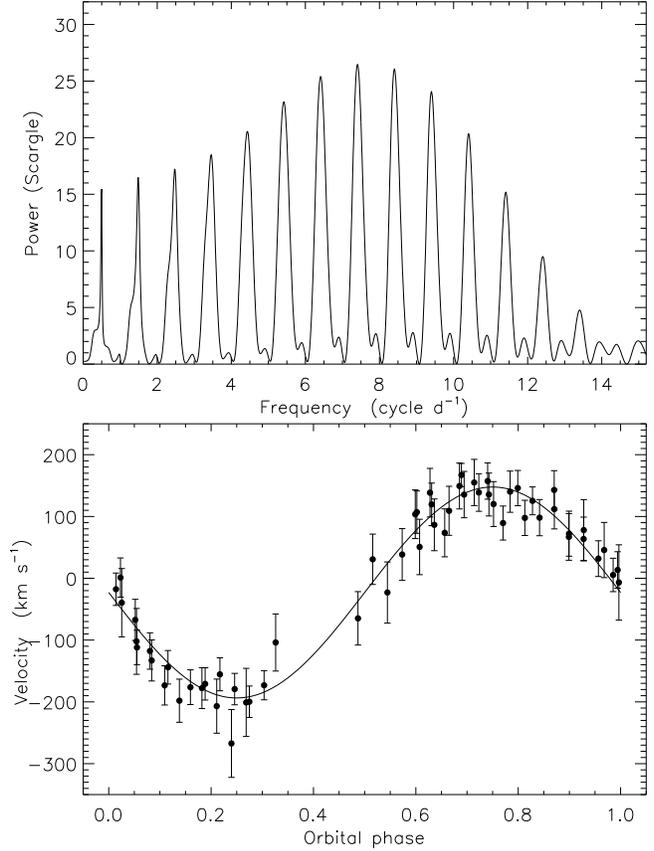}\\
\caption{\label{fig:2238:rvplot} {\it Upper panel:} Scargle periodogram of
the RVs of SDSS\,J2238 measured using a single Gaussian with FWHM 850\kms.
{\it Lower panel:} measured RVs (filled circles) compared to the best-fitting
spectroscopic orbit (solid line).} \end{figure}

SDSS\,J2238 was found to be a CV by \citet{Berg+92apjs} in the course of the Large Bright Quasar Survey \citep{Foltz+87aj}. It was given the provisional name of Aqr\,1 in the now-frozen catalogue of \citet{Downes+01pasp}. An SDSS spectrum was presented by \citet{Szkody+03aj}, along with seven intermediate-resolution spectra and 3.75 hours of unfiltered time-series photometry with exposure times of 600\,s. An orbital period of $\sim$2\,h was estimated from the spectra (although they only cover 1.75\,h), and no periodic variation was noticed in the photometry. The \ion{He}{II} 4686\,\AA\ emission in the SDSS spectrum was strong enough for \citet{Szkody+03aj} to suggest that the object may be a magnetic CV.

\citet{Woudt++04mn} presented 15\,h of unfiltered photometry of SDSS\,J2238 over five nights, obtained using the SAAO 1.9\,m telescope and UCT high-speed photometer. They found periodicities at 6.7284\,min and 193.5\,min, the second of which was found to be the `probable' orbital period. The first of these was attributed to the spin period of the white dwarf, confirming the magnetic nature of the object.

We obtained a total of 59 spectra over three consecutive nights (Table\,\ref{tab:obslog}) using the IMACS spectrograph on the Magellan Baade telescope (Section\,\ref{sec:obs:mag}). RV measurements of the H$\alpha$ line indicate an orbital period of 194\,min for all widths of Gaussian functions tried. The best results are found using a single Gaussian of width 850\kms. A Scargle periodogram of these observations is highly aliased (Fig.\,\ref{fig:2238:rvplot}) but bootstrapping simulations give a probability of 88\% that the highest peak is the correct one. Fitting a spectroscopic orbit to the RVs gives an orbital period of $\Porb = 194.30 \pm 0.16$\,min. Whilst this solution gives a scatter in the observations of $\sigma_{\rm rms} = 27.3$\kms, the alternative alias periods of 171 and 224 min have scatter of 35.0 and 34.1\kms, respectively.

Our orbital period of 194.30\,min is in good agreement with the `probable orbital period' obtained by \citet{Woudt++04mn}, which supports both our alias discrimination and their interpretation. The discrepant value of $\sim$2\,h favoured by \citet{Szkody+03aj} is probably due to the shortage of observational data available to that study. The parameters of the final spectroscopic orbit are given in Table\,\ref{tab:orbits} and phase-binned and trailed spectra around the H$\alpha$ and He\,I 6678\,\AA\ emission lines are shown in Fig.\,\ref{fig:vlt:trailed}. The 6.7\,min photometric periodicity found by \citet{Woudt++04mn} is not resolved by our spectra, which have a cadence of 5.5\,min.


\section{SUMMARY AND DISCUSSION}                                \label{sec:conclusion} \label{sec:discussion}

\begin{table} \begin{center}
\caption{\label{tab:result} Summary of the orbital
periods obtained for the objects studied in this work.}
\setlength{\tabcolsep}{4pt}
\begin{tabular}{l r@{.}l@{\,$\pm$\,}r@{.}l l} \hline
Object       & \multicolumn{4}{c}{Period (min)}       & Notes                                   \\
\hline
SDSS\,J0043  &  82&325  &  0&088                      & VLT spectroscopy                        \\
SDSS\,J0337  & \multicolumn{4}{c}{}                   & Faint, no RV motion noticed             \\
SDSS\,J1601  & \multicolumn{4}{c}{short period}       & More photometry needed                  \\
SDSS\,J1637  &  97&04   &  0&19                       & VLT spectroscopy                        \\
SDSS\,J1642  & 113&6    &  1&5                        & VLT spectroscopy                        \\
SDSS\,J1658  &  98&012  &  0&065                      & VLT spectroscopy, low state             \\
SDSS\,J1659  & \multicolumn{4}{c}{}                   & NTT photometry                          \\
SDSS\,J2232  & \multicolumn{4}{c}{}                   & VLT spec., period may be $\sim$4.5\,hr  \\
SDSS\,J2238  & 194&30   &  0&16                       & Magellan spectroscopy                   \\
\hline \end{tabular} \end{center} \end{table}

We have presented time-resolved photometry and spectroscopy of nine faint CVs which were identified by the SDSS. For five of these systems we have determined orbital periods (Table\,\ref{tab:result}), and four of these are shorter than the 2--3\,hour period gap apparent in the known population of CVs. This work brings the total number of SDSS CVs with measured orbital periods to approximately 110 of the total population of 212 objects.

From VLT spectroscopy of SDSS\,J0043 we found an orbital period of $\Porb = 82.325 \pm 0.088$\,min, placing this object close to the observed minimum period for hydrogen-rich CVs. Its spectrum shows a strong contribution from the white dwarf in the system, indicating that the accretion disc is faint and the mass transfer rate is low. We have used Doppler tomography to decompose the spectra into a Doppler map in velocity space. The map shows a circular accretion disc and an inner emission peak. The latter feature can be attributed to the irradiated inner face of the secondary star, if the velocity amplitude measured from the H$\alpha$ emission line overestimates the motion of the white dwarf by a factor of two. The Doppler map shows enhanced emission from two bright regions on the accretion disc. If the inner emission peak does indeed come from the secondary star, one of these bright regions is in the correct position to be a bright spot caused by the mass transfer stream impacting the disc. This interpretation is supported by its presence in a Doppler map of the \ion{He}{I} 6678\,\AA\ line. The second region of enhanced emission is in an unusual position in velocity space and its origin is not straightforwardly explicable. A short light curve of SDSS\,J0043 shows evidence of a variation with a period of $207 \pm 1$\,s and an amplitude of $17 \pm 5$\,mmag. The white dwarf component may be a ZZ\,Ceti-type pulsating star.

SDSS\,J1637 was observed with the VLT during quiescence, resulting in an orbital period measurement of $\Porb = 97.01 \pm 0.19$\,min. This object is a dwarf nova which has previously been observed in superoutburst, when superhumps with a period of $99.75 \pm 0.86$\,min were detected. Using the calibrations presented by \citet{Patterson98pasp} and \citet{Knigge06mn}, we find $q \approx 0.16$ from these two period measurements. Assuming a white dwarf mass of 0.8\Msun\ gives an estimated secondary mass of 0.13\Msun. This is in good agreement with the expected properties of a CV with $\Porb = 97$\,min, and together with the velocity amplitude from the H$\alpha$ emission line point to the system having an orbital inclination of roughly 40$^\circ$.

SDSS\,J1642 was studied both photometrically with the NTT and spectroscopically with the VLT. The VLT data yield an unambiguous period of $\Porb = 113.6 \pm 1.5$\,min. The NTT photometry shows a number of features and results in a periodogram with a small forest of peaks at frequencies below 15\cd. The spectroscopic period is in best agreement with the peak corresponding to a period of $110.60 \pm 0.16$\,min, but we prefer the spectroscopic value for our final orbital period measurement. Doppler maps of the H$\alpha$ and \ion{He}{I} emission lines reveal a clear accretion disc and bright spot as well as weak emission from the secondary star. More surprisingly, the \ion{He}{I} maps show an accretion disc with much higher velocities than those for H$\alpha$, indicating that the emitting region for \ion{He}{I} is physically smaller and closer to the white dwarf than for H$\alpha$. The H$\alpha$ map also shows strong emission arising from the part of the disc opposite the bright spot, but the short duration of our spectroscopic observations means this unusual feature could be caused by brightness variation which differ from orbit to orbit.

SDSS\,J1658 is perhaps the most unusual system studied in this work. Its SDSS spectrum shows the strong emission lines characteristic of a short-period CV whose light is dominated by a hydrogen-rich accretion disc. However, our VLT spectra show a much weaker and narrower central emission line, and broad absorption features arising from both the white dwarf and secondary star, but no flux which could be unambiguously assigned to an accretion disc. The object was only 0.4\,mag fainter during our observations than when the SDSS spectrum was taken, so the strong emission lines in that spectrum were accompanied by only weak continuum flux from the accretion disc. The velocity variation of the narrow H$\alpha$ emission in our VLT spectra yields a period measurement of $\Porb = 98.012 \pm 0.065$\,min, confirming that this object is a short-period binary star system. The velocity amplitude (125\kms) is too large for the white dwarf, \reff{so we have attempted to assign it to the secondary star. The observational constraints can be satsified in this scenario if the white dwarf is massive ($\la$1\Msun) and the orbital inclination is low (10--30$^\circ$).}

A modest number of photometric and spectroscopic observations of SDSS\,J2232 suggest that this is a dwarf nova with an orbital period close to 4.5\,hr. \reff{From the flux contribution of the secondary star we infer a distance of roughly 3\,kpc, corresponding to distance of 2\,pkc from the Galactic plane.} Further investigation is needed to prove its membership of this old stellar population.

We obtained 59 Magellan spectra of SDSS\,J2238, which is the brightest of the five objects for which we determine orbital periods in this work. Velocity measurements of its H$\alpha$ emission line give a period of $\Porb = 194.30 \pm 0.16$. This value is in good agreement with a published photometric period, which also confirmed that the system contains a magnetic white dwarf with a rotational period of 6.7284\,min.

These observations provide further confirmation that the faintest of the CVs identified by the SDSS have predominantly short orbital periods. Theoretical population studies of CVs predict a huge population of faint short-period CVs which has not previously been detected. Our observations are now uncovering this `quiet majority' of the CV population. The unusual characteristics of several of the objects studied in this work show that even the short-period CVs demonstrate impressively varied behaviour, many aspects of which cannot easily be explained in the standard picture of CV structure. Our project to study the SDSS CV population is invaluable for extending our knowledge of these fascinating objects.


\section*{ACKNOWLEDGEMENTS}

The reduced spectra and radial velocity measurements presented in this work will be made available at the CDS ({\tt http://cdsweb.u-strasbg.fr/}) and at {\tt http://www.astro.keele.ac.uk/$\sim$jkt/}. Based on observations made with ESO Telescopes at the La Silla and Paranal Observatories under programme ID 079.D-0024. Some data presented here have been taken using ALFOSC, which is owned by the Instituto de Astrofis\'{\i}ca de Andaluc\'{\i}a (IAA) and operated at the Nordic Optical Telescope under agreement between IAA and the NBIfAFG of the Astronomical Observatory of Copenhagen.

We would like to acknowledge the referee, John Thorstensen, for an insightful report. JS and CMC acknowledge financial support from PPARC in the form of a postdoctoral research assistant position. DS acknowledges an STFC Advanced Fellowship. The following internet-based resources were used in research for this paper: the ESO Digitised Sky Survey; the NASA Astrophysics Data System; the SIMBAD database operated at CDS, Strasbourg, France; and the ar$\chi$iv scientific paper preprint service operated by Cornell University.

Funding for the Sloan Digital Sky Survey (SDSS) has been provided by the Alfred P.\ Sloan Foundation, the Participating Institutions, the National Aeronautics and Space Administration, the National Science Foundation, the U.\ S.\ Department of Energy, the Japanese Monbukagakusho, and the Max Planck Society. The SDSS website is {\tt http://www.sdss.org/}.


\bibliographystyle{mn_new}

\label{lastpage}

\end{document}